\documentclass[aps,pra,reprint,floatfix,showpacs]{revtex4-1}
\usepackage{array}
\usepackage{multirow}
\usepackage{amsmath}
\usepackage{graphicx}

\newcommand*{\pr}[1]{\mathcal{#1}}
\newcommand*{\refneq}[1]{(\ref{#1})}
\newcommand*{\refeq}[1]{Eq.\ (\ref{#1})}

\begin{document}

\title{Sum rules for spin-$1/2$ quantum gases in states with well-defined
spins: spin-independent interactions and spin-dependent external fields.}

\author{Vladimir A. Yurovsky}

\affiliation{School of Chemistry, Tel Aviv University, 6997801 Tel Aviv, Israel}

\date{\today}
\begin{abstract}
Analytical expressions are derived for sums of matrix elements and
their squared moduli over many-body states with given total spin ---
the states built from spin and spatial wavefunctions belonging to
multidimensional irreducible representations of the symmetric group,
unless the total spin has the maximal allowed value. For spin-dependent
one-body interactions with external fields and spin-independent two-body
ones between the particles, the sum dependence on the many-body states
is given by universal factors, which are independent of the interaction
details and Hamiltonians of non-interacting particles. The sum rules
are applied to perturbative analysis of energy spectra. 
\end{abstract}

\pacs{67.85.Fg,67.85.Lm,02.20.-a,03.65.Fd}

\maketitle

\section*{Introduction}

Calculations of quantum-mechanical system properties require matrix
elements between its states. For complex systems, even a calculation
of the matrix elements can constitute a complicated problem. However,
in certain cases, sum rules can be derived from general principles,
providing analytical expressions for sums of matrix elements or their
products. The Thomas-Reiche-Kuhn and the Bethe sum rules were obtained
at early years of quantum mechanics. These and similar rules (see
\cite{bethe}) are formulated for weighted sums of oscillator strengths,
which are proportional to squared moduli of transition matrix elements,
over certain sets of eigenstates. The rules were applied to radiative
transitions and scattering problems. Sum rules for dynamic structure
factors (see \cite{pitaevskii}) are employed to obtain information
on collective behavior of many-body systems. Various sum rules are
also used in nuclear and solid-state physics, as well as in quantum
field theory.

The present work derives sum rules for many-body systems of indistinguishable
spinor particles. The particles can be composite, e.g., atoms or molecules,
and the spin can be either a real angular momentum of the particle
or a formal spin, whose projections are attributed to the particle's
internal states (e.g. hyperfine states of atoms). In the latter case,
the particle spin $\frac{1}{2}$ means that only two internal states
are present in the system. This formal spin is not related to the
real, physical, spin of the particles, which can be either bosons
or fermions. 

Many-body states of spinor particles can be described in two ways
(see \cite{landau}). In the first one, each particle is characterized
by its spin projection and coordinate, and the total wavefunction
is symmetrized for bosons or antisymmetrized for fermions over permutations
of all particles {[}see \refeq{tilPsinsig} in Sec. \ref{secWavFun}
below{]}. The second approach is based on collective spin and spatial
wavefunctions. These wavefunctions depend on spins or coordinates,
respectively, of all particles and form representations of the symmetric
group (see \cite{hamermesh,elliott,kaplan,pauncz_symmetric}). In
the case of spin-$\frac{1}{2}$ particles, the representation is unambiguously
related to the total many-body spin. If the total spin is less than
the maximal allowed one ($N/2$ for $N$ particles), the wavefunctions
belong to multidimensional, non-Abelian, irreducible representations
of the symmetric group (see \cite{hamermesh,elliott,kaplan,pauncz_symmetric}),
beyond the conventional paradigm of symmetric-antisymmetric functions.
The symmetric or antisymmetric total wavefunction --- such functions
only are allowed by the Pauli exclusion principle --- is represented
as a sum of products of the spin and spatial wavefunctions {[}see
\refeq{PsiSnl} in Sec. \ref{secWavFun} below{]}.

In the case of non-interacting particles in spin-independent potentials,
all states with the given set of spatial quantum numbers are energy-degenerate
and the two kinds of wavefunctions are applicable, related by a linear
transformation. The effect of spin-independent interactions between
particles was analyzed by Heitler \cite{heitler1927} using the theory
of the symmetric group irreducible representations. That work demonstrates
that the average energy of states within given irreducible representation
is proportional to a certain sum of the representation characters.
The character dependence on the representation lifts the degeneracy
of states related to different representations, and the wavefunctions
with defined individual spin projections become inapplicable. It is
a generalization of the well-known energy splitting between the singlet
and triplet states in two-electron problems.

Although the derivation \cite{heitler1927}, being done at early years
of quantum mechanics, did not take into account spin degrees of freedom
and supposed that total wavefunctions can have arbitrary permutation
symmetry, the results remain valid for symmetric or antisymmetric
total wavefunctions, composed from spin and spatial functions of arbitrary
symmetry. Matrix elements of spin-independent Hamiltonians between
the latter wavefunctions can be reduced to the matrix elements between
spatial wavefunctions due to orthogonality of the spin wavefunctions
(see Sec. \ref{secTwoBody}). Besides justification of the Heitler
results, this reduction provides basis for spin-free quantum chemistry
(see \cite{kaplan,pauncz_symmetric}) --- the method of calculations
of energies and other properties of atoms and molecules. 

Spinor quantum gases are intensively studied starting from the first
experimental \cite{myatt1997,stamper1998} and theoretical \cite{ho1998,ohmi1998}
works (see book \cite{pitaevskii}, reviews \cite{stamper2013,guan2013}
and references therein). The collective spin and spatial wavefunctions
were used in derivation of exact quantum solutions for one-dimensional
homogeneous gas \cite{yang1967,sutherland1968} and in analyses of
selection rules and correlations \cite{yurovsky2014}. $SU(M)$-symmetric
gases, introduced in Refs. \cite{honerkamp2004,gorshkov2010,cazalilla2009}
and recently observed in Refs. \cite{zhang2014,scazza2014}, are described
in similar way \cite{gorshkov2010}, where the total wavefunction
is composed of spin and electronic functions. 

Other forms of many-body wavefunctions with defined total spin have
been employed as well. The Lieb-Mattis theorem for ordering of energy
levels in fermionic systems has been derived in Ref. \cite{lieb1962}.
One-dimensional gas of spin-$\frac{1}{2}$ fermions in arbitrary potential
has been analyzed for hardcore zero-range interactions in Ref. \cite{guan2009},
where an exact solution was derived, and for zero-range interactions
of arbitrary strength in Ref. \cite{yang2009}, where qualitative
properties of energy spectra are presented. An exact solution for
one-dimensional hardcore Bose-Fermi mixture was derived in Ref. \cite{fang2011}.
Intersystem degeneracies in spin-$\frac{1}{2}$ Fermi gases and energy
spectra for certain few-body systems have been obtained in Ref. \cite{daily2012}.
Symmetries of trapped and interacting bosons and fermions and qualitative
behavior of the energy spectra at intermediate interaction strengths
were analyzed in Refs. \cite{harshman2014,harshman2015}.

The sum of matrix elements of spin-independent interparticle interactions
directly follows from the Heitler results \cite{heitler1927}. The
present paper provides the sums of squared moduli of these matrix
elements, as well as sums of matrix elements and their squared moduli
for spin-dependent external fields. Such fields can be used for transfer
of population between states with different total spins, as described
in \cite{yurovsky2014}. Besides, spin-changing matrix elements can
provide an estimate of stability of the well-defined-spin states. 

Section \ref{secWavFun} sets the analyzed problem and provides representations
of spin, spatial, and total wavefunctions for separable spin and spatial
degrees of freedom and for non-interacting particles. Wavefunctions
with defined particle spin projections are discussed in this section
too. Section \ref{secSumRules} contains derivation of the sum rules.
Matrix elements of spin-dependent external fields for different total
spin projections are related using the Wigner-Eckart theorem. Then
sums of these matrix elements and their squared moduli are calculated
for the maximal allowed spin projections. Sum rules for spin-independent
interactions between particles are provided in Sec. \ref{secTwoBody}.
The sum rules are applied to description of the shifts and splittings
of energy levels in Sec. \ref{sec_energies}. The quantitative properties
of energy spectra are provided for arbitrary number of particles in
the regime of weak interactions using perturbation theory. Appendix
contains calculation of sums, used in Sec. \ref{secSumRules}.

\section{The Hamiltonian and wavefunctions\label{secWavFun}}

Consider a system of $N$ particles with the Hamiltonian 
\begin{equation}
\hat{H}=\hat{H}_{\mathrm{spat}}+\hat{H}_{\mathrm{spin}}\label{HSpinSpat}
\end{equation}
being a sum of the spin-independent $\hat{H}_{\mathrm{spat}}$ and
coordinate-independent $\hat{H}_{\mathrm{spin}}$. Each of $\hat{H}_{\mathrm{spat}}$
and $\hat{H}_{\mathrm{spin}}$ is permutation-invariant. 

The total wavefunction is expressed in the form 
\begin{equation}
\Psi_{nl}^{(S)}=f_{S}^{-1/2}\sum_{t}\Phi_{tn}^{(S)}\Xi_{tl}^{(S)}.\label{PsiSnl}
\end{equation}
Here spatial $\Phi_{tn}^{(S)}$ and spin $\Xi_{tl}^{(S)}$ functions
form bases of irreducible representations of the symmetric group $\pr S_{N}$
of $N$-symbol permutations\cite{hamermesh,elliott,kaplan,pauncz_symmetric}.
This means that a permutation $\pr P$ of the particles transforms
each function to a linear combination of functions in the same representation,
\begin{eqnarray*}
\pr P\Phi_{tn}^{(S)} & = & \mathrm{sgn}(\pr P)\sum_{t'}D_{t't}^{[\lambda]}(\pr P)\Phi_{t'n}^{(S)}\\
\pr P\Xi_{tl}^{(S)} & = & \sum_{t'}D_{t't}^{[\lambda]}(\pr P)\Xi_{t'l}^{(S)}
\end{eqnarray*}
Here the factor $\mathrm{sgn}(\pr P)$ is the permutation parity for
fermions and $\mathrm{sgn}(\pr P)\equiv1$ for bosons. This factor
provides the proper permutation symmetry of the total wavefunction
\begin{equation}
\pr P\Psi_{nl}^{(S)}=\mathrm{sgn}(\pr P)\Psi_{nl}^{(S)}\label{ExclPrinc}
\end{equation}

The matrices of the Young orthogonal representation \cite{hamermesh,elliott,kaplan,pauncz_symmetric}
$D_{t't}^{[\lambda]}(\pr P)$ of the symmetric group $\pr S_{N}$
are associated with the two-row Young diagrams $\lambda=[N/2+S,N/2-S]$,
which are unambiguously related to the total spin $S$. Different
representations, associated with the same Young diagram, are labeled
by multi-indices $n$ and $l$ for the spatial and spin functions,
respectively. The representation basis functions are labeled by the
standard Young tableaux $t$ and $t'$ of the shape $\lambda$. The
dimension of the representation is equal to the number of different
tableaux, 
\begin{equation}
f_{S}=\frac{N!(2S+1)}{(N/2+S+1)!(N/2-S)!}.\label{flambda}
\end{equation}
If $S=N/2$, $f_{S}=1$, $D_{t't}^{[\lambda]}(\pr P)=1$, and the
functions $\Phi_{tn}^{(S)}$ and $\Xi_{tl}^{(S)}$ remain unchanged
on permutations of particles or change their sign ($\Phi_{tn}^{(S)}$
for fermions). Otherwise, the functions belong to multidimensional,
non-Abelian irreducible representations of the symmetric group. For
example, the states of $N=3$ particles with $S=1/2$ are associated
with the Young diagram $[2,1]$ and there are $f_{S}=2$ standard
Young tableaux with the Yamanouchi symbols (see \cite{hamermesh,elliott})
$(2,1,1)$ and $(1,2,1)$. 

The Young orthogonal matrices obey the orthogonality relation\cite{kaplan,pauncz_symmetric}
\begin{equation}
\sum_{\pr P}D_{t'r'}^{[\lambda']}(\pr P)D_{tr}^{[\lambda]}(\pr P)=\frac{N!}{f_{S}}\delta_{tt'}\delta_{rr'}\delta_{\lambda\lambda'},\label{OrthRel}
\end{equation}
the general relation for representation matrices 
\begin{equation}
\sum_{t}D_{r't}^{[\lambda]}(\pr P)D_{tr}^{[\lambda]}(\pr Q)=D_{r'r}^{[\lambda]}(\pr P\pr Q),\label{RepProd}
\end{equation}
and the relation for orthogonal matrices 
\begin{equation}
D_{tr}^{[\lambda]}(\pr P^{-1})=D_{rt}^{[\lambda]}(\pr P).\label{InvOrthMat}
\end{equation}

Additional relations can be obtained for elements of the first column
$D_{t[0]}^{[\lambda]}(\pr P)$ of the Young orthogonal matrices. Here
$[0]$ is the first Young tableau, in which the symbols are arranged
by rows in the sequence of natural numbers. For example, the Young
tableaux $[0]$ have Yamanouchi symbols $(2,1,1)$, $(2,2,1,1)$,
and $(2,1,1,1)$ for the Young diagrams $[2,1]$, $[2^{2}]$, and
$[3,1]$, respectively. Each permutation involving symbols between
$j_{\mathrm{min}}$ and $j_{\mathrm{max}}$ can be written as a product
of elementary transpositions $\pr P_{jj+1}$ with $j_{\mathrm{min}}\leq j<j_{\mathrm{max}}$
(see \cite{elliott,kaplan,pauncz_symmetric}). According to the Young
orthogonal matrix calculation rules (see \cite{elliott,kaplan,pauncz_symmetric}),
$D_{rt}^{[\lambda]}(\pr P_{jj+1})=\delta_{rt}$ if $j$ and $j+1$
are in the same row of the Young tableau $t$. Then \refeq{RepProd}
leads to $D_{t[0]}^{[\lambda]}(\pr P)=\delta_{t[0]}$ if the permutation
$\pr P$ involves the symbols in one row only and can be, therefore,
written as a product of elementary transpositions of symbols in the
same row. Let $\pr P'$ and $\pr P''$ be, respectively, arbitrary
permutations of the symbols in the first and in the second row of
the Young tableau $[0]$, which do not permute symbols between the
rows. Then we get, using \refeq{RepProd}, 
\begin{equation}
D_{t[0]}^{[\lambda]}(\pr P\pr P'\pr P'')=\sum_{r}D_{tr}^{[\lambda]}(\pr P)D_{r[0]}^{[\lambda']}(\pr P'\pr P'')=D_{t[0]}^{[\lambda]}(\pr P).\label{DPpPpp}
\end{equation}

The spatial and spin wavefunctions form orthonormal basis sets,
\begin{align}
\langle\Phi_{t'n'}^{(S')}|\Phi_{tn}^{(S)}\rangle & =\delta_{S'S}\delta_{t't}\delta_{n'n}\\
\langle\Xi_{t'l'}^{(S')}|\Xi_{tl}^{(S)}\rangle & =\delta_{S'S}\delta_{t't}\delta_{l'l}\label{Xiorth}
\end{align}

The spatial functions of non-interacting particles are expressed as
\cite{kaplan,pauncz_symmetric} 
\begin{equation}
\tilde{\Phi}_{tr\{n\}}^{(S)}=\left(\frac{f_{S}}{N!}\right)^{1/2}\sum_{\pr P}\mathrm{sgn}(\pr P)D_{tr}^{[\lambda]}(\pr P)\prod_{j=1}^{N}\varphi_{n_{j}}(\mathbf{r}_{\pr Pj})\label{tilPhi}
\end{equation}
in terms of the spatial orbitals --- the eigenfunctions $\varphi_{n}(\mathbf{r})$
of the one-body Hamiltonian of non-interacting particle $\hat{H}_{0}(j)$,
\begin{equation}
\hat{H}_{0}(j)\varphi_{n}(\mathbf{r}_{j})=\varepsilon_{n}\varphi_{n}(\mathbf{r}_{j}),\label{SchrH0}
\end{equation}
where $\mathbf{r}_{j}$ is the $D$-dimensional coordinate of $j$th
particle ($D$ can be either $1$, $2$, or $3$ in real physical
systems). The representation is determined by the set of the spatial
quantum numbers $\{n\}$ and the Young tableau $r$, which can take
one of $f_{S}$ values. Then the multi-index $n$ is specifically
chosen as $r\{n\}$. All quantum numbers $n_{j}$ in the set $\{n\}$
are supposed to be different. This situation takes place in non-degenerate
gases, when probabilities of multiple occupation of spatial states
are negligibly small, although the multiple occupation is not forbidden
by itself. Another example is an optical lattice in the unit-filling
regime \cite{yurovsky2014}. 

The functions \refneq{tilPhi} satisfy the Schr\"odinger equation
\[
\sum_{j=1}^{N}\hat{H}_{0}(j)\tilde{\Phi}_{tr\{n\}}^{(S)}=\sum_{j=1}^{N}\varepsilon_{n_{j}}\tilde{\Phi}_{tr\{n\}}^{(S)}
\]
Their eigenenergies are independent of $r$. Therefore, there are
$f_{S}$ degenerate states of non-interacting particles for each set
$\{n\}$. Tilde denotes wavefunctions corresponding to the spatial
Hamiltonian without interactions between particles. Then Eq. \refneq{PsiSnl}
gives us the total wavefunctions of particles with no coordinate-dependent
interactions 
\begin{equation}
\tilde{\Psi}_{r\{n\}l}^{(S)}=f_{S}^{-1/2}\sum_{t}\tilde{\Phi}_{tr\{n\}}^{(S)}\Xi_{tl}^{(S)}.\label{tilPsiSrnl}
\end{equation}

In the absence of interactions between spins, the spin wavefunction
are eigenfunctions of the total spin projection operator $\hat{S}_{z}$
and can be expressed as 
\begin{equation}
\Xi_{tS_{z}}^{(S)}=C_{SS_{z}}\sum_{\pr P}D_{t[0]}^{[\lambda]}(\pr P)\prod_{j=1}^{N/2+S_{z}}|\uparrow(\pr Pj)\rangle\prod_{j=N/2+S_{z}+1}^{N}|\downarrow(\pr Pj)\rangle.\label{XiStSz}
\end{equation}
Here the multi-index $l$ is specifically chosen as the total spin
projection $S_{z}$. In the case of the spin wavefunction, each of
two spin states, $|\uparrow\rangle$ and $|\downarrow\rangle$, has
to be occupied by several particles, if $N>2$. Hence the normalization
factor \cite{yurovsky2013}

\begin{equation}
C_{SS_{z}}=\frac{1}{(N/2+S_{z})!(N/2-S)!}\sqrt{\frac{(2S+1)(S+S_{z})!}{(N/2+S+1)(2S)!(S-S_{z})!}}\label{CSSz}
\end{equation}
differs from the one in the spatial wavefunction \refneq{tilPhi}.
Besides, the Young tableau $r$ can take now only the value of $[0]$.
As a result, only one representation is associated with given total
spin $S$ and its projection $S_{z}$. The total wavefunction with
the defined $S_{z}$ is then expressed as

\begin{equation}
\Psi_{nS_{z}}^{(S)}=f_{S}^{-1/2}\sum_{t}\Phi_{tn}^{(S)}\Xi_{tS_{z}}^{(S)}.\label{PsiSnSz}
\end{equation}

In combination with the spatial wavefunction \refneq{tilPhi}, the
spin wavefunctions lead to the total wavefunctions of non-interacting
particles,

\begin{equation}
\tilde{\Psi}_{r\{n\}S_{z}}^{(S)}=f_{S}^{-1/2}\sum_{t}\tilde{\Phi}_{tr\{n\}}^{(S)}\Xi_{tS_{z}}^{(S)}\label{tilPsiSrnSz}
\end{equation}
(again, tilde denotes that the wavefunctions involve spatial orbitals
$\varphi_{n}(\mathbf{r})$ of non-interacting particles). There are
$f_{S}$ wavefunctions, labeled by the Young tableau $r$, having
the total spin $S$ and the set of spatial quantum numbers $\{n\}$.
Then the total number of wavefunctions with the given total spin projection
$S_{z}$ will be 
\begin{equation}
\sum_{S=S_{z}}^{N/2}f_{S}=N!/[(N/2+S_{z})!(N/2-S_{z})!].\label{sumfS}
\end{equation}

In the alternative approach, mentioned in Introduction, each particle
has a given spin projection and the total many-body wavefunction is
represented as (see \cite{landau}) 
\begin{equation}
\tilde{\Psi}_{\{n\}\{\sigma\}}=(N!)^{-1/2}\sum_{\pr P}\mathrm{sgn}(\pr P)\prod_{j=1}^{N}\varphi_{n_{j}}(\mathbf{r}_{\pr Pj})|\sigma_{j}(\pr Pj)\rangle,\label{tilPsinsig}
\end{equation}
where the spin projection $\sigma_{j}$ can be either $\uparrow$
or $\downarrow$ and given total spin projection $S_{z}$, the set
$\{\sigma\}$ contains $N/2+S_{z}$ spins $\uparrow$ and $N/2-S_{z}$
spins $\downarrow$. For a fixed set of spatial quantum numbers $\{n\}$,
the number of such states is the number of distinct choices of $N/2+S_{z}$
particles with spin up, and is then equal to the number \refneq{sumfS}
of the states \refneq{tilPsiSrnSz}. Then the sets of degenerate states
$\tilde{\Psi}_{r\{n\}S_{z}}^{(S)}$ and $\tilde{\Psi}_{\{n\}\{\sigma\}}$
can be related by an unitarily transformation. For interacting particles,
the energy degeneracy of states $\tilde{\Psi}_{r\{n\}S_{z}}^{(S)}$
is lifted, as shown by Heitler \cite{heitler1927} and will be discussed
in Sec. \ref{sec_energies}, and such transformation becomes impossible.

\section{Sum rules for one-body interactions\label{secSumRules}}

\subsection{The spin-projection dependence}

Permutation-invariant interactions of particles with external fields
can be expressed in terms of the spherical scalar
\begin{equation}
\hat{U}=\sum_{j}U(\mathbf{r}_{j})\label{Uscal}
\end{equation}
and three spherical vector components
\begin{equation}
\hat{U}_{0}=\sum_{j}U(\mathbf{r}_{j})\hat{s}_{z}(j),\quad\hat{U}_{\pm1}=\mp\frac{1}{\sqrt{2}}\sum_{j}U(\mathbf{r}_{j})\hat{s}_{\pm}(j)\label{Uvect}
\end{equation}
(see \cite{cook1971}). Here 
\[
\hat{s}_{z}(j)=\frac{1}{2}(|\uparrow(j)\rangle\langle\uparrow(j)|-|\downarrow(j)\rangle\langle\downarrow(j)|)
\]
 is the $z$-component of the spin and 
\[
\hat{s}_{+}(j)=|\uparrow(j)\rangle\langle\downarrow(j)|,\quad\hat{s}_{-}(j)=|\downarrow(j)\rangle\langle\uparrow(j)|
\]
 are the spin raising and lowering operators for $j$th particle.
The interaction $\hat{U}_{0}$ conserves the $z$-projection of the
total many-body spin, while $\hat{U}_{\pm1}$ raises or lowers it.
The interaction of the spin-up or spin-down state can be expressed
in terms of $\hat{U}_{0}$ and the scalar $\hat{U}$,
\begin{equation}
\begin{aligned}\hat{U}_{\uparrow} & \equiv\sum_{j}U(\mathbf{r}_{j})|\uparrow(j)\rangle\langle\uparrow(j)|=\hat{U}_{0}+\frac{1}{2}\hat{U}\\
\hat{U}_{\downarrow} & \equiv\sum_{j}U(\mathbf{r}_{j})|\downarrow(j)\rangle\langle\downarrow(j)|=-\hat{U}_{0}+\frac{1}{2}\hat{U}.
\end{aligned}
\label{Uupdown}
\end{equation}

\begin{table}
\protect\caption{Coefficients $X_{S_{z}k}^{(S,S',1)}$ in  \refeq{UvectWE} }

\begin{tabular}{|c|c|c|}
\hline 
\multirow{2}{*}{$k$} & \multicolumn{2}{c|}{$S-S'$}\tabularnewline
\cline{2-3} 
 & $0$ & $1$\tabularnewline
\hline 
-1 & $\frac{\sqrt{(S-S_{z}+1)(S+S_{z})}}{\sqrt{2}S}$ & $\sqrt{\frac{(S+S_{z}-1)(S+S_{z})}{2S(2S-1)}}$\tabularnewline
\hline 
0 & $\frac{S_{z}}{S}$ & $-\sqrt{\frac{S^{2}-S_{z}^{2}}{S(2S-1)}}$\tabularnewline
\hline 
1 & $-\frac{\sqrt{(S-S_{z})(S+S_{z}+1)}}{\sqrt{2}S}$ & $\sqrt{\frac{(S-S_{z}-1)(S-S_{z})}{2S(2S-1)}}$\tabularnewline
\hline 
\end{tabular}\label{TabUvec}
\end{table}

Consider matrix elements of the spherical vector and scalar interactions
between eigenfunctions \refneq{PsiSnSz} of $\hat{S}_{z}$. Their
dependence on $S_{z}$ follows from the Wigner-Eckart theorem (see
\cite{edmonds}). The matrix elements of the spherical scalar \refneq{Uscal}
are diagonal in spins and independent of the spin projection,
\[
\langle\Psi_{n'S'_{z}}^{(S')}|\hat{U}|\Psi_{nS_{z}}^{(S)}\rangle=\delta_{SS'}\delta_{S_{z}S'_{z}}\langle\Psi_{n'S}^{(S)}|\hat{U}|\Psi_{nS}^{(S)}\rangle.
\]
According to the Wigner-Eckart theorem, the matrix elements of the
spherical vector components \refneq{Uvect} can be factorized into
the $3j$-Wigner symbols and the reduced matrix elements
\[
\langle\Psi_{n'S'_{z}}^{(S')}|\hat{U}_{k}|\Psi_{nS_{z}}^{(S)}\rangle=(-1)^{S'-S'_{z}}\left(\begin{array}{ccc}
S' & 1 & S\\
-S'_{z} & k & S_{z}
\end{array}\right)\langle n',S'||\hat{U}||n,S\rangle.
\]
Then the reduced matrix elements are expressed in terms of the matrix
elements of $\hat{U}_{k}$ for the maximal allowed spin projection
\[
\langle n',S'||\hat{U}||n,S\rangle=\left(\begin{array}{ccc}
S' & 1 & S\\
-S' & S'-S & S
\end{array}\right)^{-1}\langle\Psi_{n'S'}^{(S')}|\hat{U}_{S'-S}|\Psi_{nS}^{(S)}\rangle,
\]
and the matrix elements with arbitrary spin projections can be expressed
as
\begin{equation}
\langle\Psi_{n'S'_{z}}^{(S')}|\hat{U}_{k}|\Psi_{nS_{z}}^{(S)}\rangle=\delta_{S'_{z}S_{z}+k}X_{S_{z}k}^{(S,S',1)}\langle\Psi_{n'S'}^{(S')}|\hat{U}_{S'-S}|\Psi_{nS}^{(S)}\rangle\label{UvectWE}
\end{equation}
with the factors
\begin{multline*}
X_{S_{z}k}^{(S,S',q)}=(-1)^{S'-S_{z}-k}\left(\begin{array}{ccc}
S & S' & q\\
S_{z} & -S_{z}-k & k
\end{array}\right)\\
\times\left(\begin{array}{ccc}
S & S' & q\\
S & -S' & S'-S
\end{array}\right)^{-1}
\end{multline*}
 Here $S'\leq S$ and, according to the properties of the $3j$-Wigner
symbols, the matrix elements \refneq{UvectWE} vanish if $|S-S'|>1$
(in agreement to the selection rules \cite{yurovsky2014}). Values
of non-vanishing coefficients, calculated with the $3j$-Wigner symbols
\cite{landau,edmonds}, are presented in Tab. \ref{TabUvec}. Hermitian
conjugate of \refeq{UvectWE}, together with relations $\hat{U}_{+1}=-\hat{U}_{-1}^{\dagger}$
and $\hat{U}_{0}=\hat{U}_{0}^{\dagger}$, gives us the matrix elements
for $S'=S+1$.

Thus, each matrix element of a spin-dependent one-body interaction
with an external field is related to matrix elements for the maximal
allowed spin projections, which will be evaluated in the next section.

\subsection{Matrix elements for non-interacting particles}

Matrix elements of the spherical scalar \refneq{Uscal} can be evaluated
exactly for general spin wavefunctions. Due to the orthogonality of
the spin wavefunctions \refneq{Xiorth}, the matrix elements are diagonal
in spin quantum numbers and can be reduced to the matrix elements
between spatial wavefunctions,
\begin{multline}
\langle\tilde{\Psi}_{r'\{n'\}l'}^{(S')}|\hat{U}|\tilde{\Psi}_{r\{n\}l}^{(S)}\rangle=\delta_{SS'}\delta_{ll'}\frac{1}{f_{S}}\\
\times\sum_{t}\sum_{i}\langle\tilde{\Phi}_{tr'\{n'\}}^{(S)}|U(\mathbf{r}_{i})|\tilde{\Phi}_{tr\{n\}}^{(S)}\rangle.\label{UtilPsitilPhi}
\end{multline}
Let us calculate the spatial matrix element for the general case,
$S\neq S'$, having in mind further analysis of spherical vectors.
Equations \refneq{tilPhi} and \refneq{InvOrthMat} lead to
\begin{multline*}
\langle\tilde{\Phi}_{t'r'\{n'\}}^{(S')}|U(\mathbf{r}_{i})|\tilde{\Phi}_{tr\{n\}}^{(S)}\rangle=\frac{\sqrt{f_{S}f_{S'}}}{N!}\sum_{\pr R,\pr Q}\mathrm{sgn}(\pr Q)D_{r't'}^{[\lambda']}(\pr Q)\\
\times\mathrm{sgn}(\pr R)D_{rt}^{[\lambda]}(\pr R)\langle\varphi_{n'_{\pr Qi}}|U(\mathbf{r}_{i})|\varphi_{n_{\pr Ri}}\rangle\prod_{i'\neq i}\delta_{n'_{\pr Qi'},n_{\pr Ri'}}.
\end{multline*}
 The Kronecker $\delta$-symbols appear here due to the orthogonality
of the spatial orbitals $\varphi_{n}$ and the absence of equal quantum
numbers in each of the sets $\{n\}$ and $\{n'\}$. Due to the $\delta$-symbols,
all but one spatial quantum numbers remain unchanging. Supposing that
the unchanged $n_{j'}$ are in the same positions in the sets $\{n\}$
and $\{n'\}$, one can see that the Kronecker symbols lead to $\pr Q=\pr R$,
and, therefore,
\begin{multline}
\langle\tilde{\Phi}_{t'r'\{n'\}}^{(S')}|U(\mathbf{r}_{i})|\tilde{\Phi}_{tr\{n\}}^{(S)}\rangle=\frac{\sqrt{f_{S}f_{S'}}}{N!}\sum_{\pr R}D_{r't'}^{[\lambda']}(\pr R)D_{rt}^{[\lambda]}(\pr R)\\
\times\langle n'_{\pr Ri}|U|n_{\pr Ri}\rangle\prod_{j'\neq\pr Ri}\delta_{n'_{j'},n_{j'}},\label{UuptilPhi}
\end{multline}
where $\langle n'|U|n\rangle=\int d^{D}r\varphi_{n'}^{*}(\mathbf{r})U(\mathbf{r})\varphi_{n}(\mathbf{r})$.
Then, substituting this expression into \refneq{UtilPsitilPhi}, using
\refneq{RepProd}, \refneq{InvOrthMat}, and the property of representations
$D_{r'r}^{[\lambda]}(\pr E)=\delta_{r'r}$, where $\pr E$ is the
identity permutation, one finally gets
\begin{equation}
\langle\tilde{\Psi}_{r'\{n'\}l'}^{(S')}|\hat{U}|\tilde{\Psi}_{r\{n\}l}^{(S)}\rangle=\delta_{SS'}\delta_{ll'}\delta_{r'r}\sum_{j=1}^{N}\langle n'_{j}|U|n_{j}\rangle\prod_{j'\neq j}\delta_{n'_{j'},n_{j'}},\label{UtilPsi}
\end{equation}
It is a special case of the matrix elements obtained by Heitler \cite{heitler1927}
and Kaplan \cite{kaplan}.

For the spherical vector interactions \refneq{Uvect}, the matrix
elements cannot be represented in so simple a form. However, rather
simple expressions can be derived for sums and sums of squared moduli
of the matrix elements between eigenfunctions of $\hat{S}_{z}$. It
is enough to consider matrix elements of $\hat{U}_{-1}$ and the spin-up
state interaction $\hat{U}_{\uparrow}$ for the maximal allowed spin
projection, $S'_{z}=S'$, $S_{z}=S$, as \refeq{Uupdown} and the
Wigner-Eckart theorem \refneq{UvectWE} relate to them each matrix
element of each interaction. In the basis of the non-interacting particle
wavefunctions \refneq{tilPsiSrnSz}, the matrix elements of $\hat{U}_{\uparrow}$
can be decomposed into the spatial and spin parts,
\begin{multline}
\langle\tilde{\Psi}_{r'\{n'\}S'}^{(S')}|\hat{U}_{\uparrow}|\tilde{\Psi}_{r\{n\}S}^{(S)}\rangle=(f_{S}f_{S'})^{-1/2}\\
\times\sum_{t,t',i}\langle\tilde{\Phi}_{t'r'\{n'\}}^{(S')}|U(\mathbf{r}_{i})|\tilde{\Phi}_{tr\{n\}}^{(S)}\rangle\langle\Xi_{t'S'}^{(S')}|\uparrow(i)\rangle\langle\uparrow(i)|\Xi_{tS}^{(S)}\rangle.\label{UuptilPsi}
\end{multline}
 The spatial matrix elements are given by \refeq{UuptilPhi}. The
spin matrix elements include projections of the spin wavefunctions
\refneq{XiStSz} 
\begin{multline*}
\langle\uparrow(i)|\Xi_{tS}^{(S)}\rangle=C_{SS}\sum_{\pr P}D_{t[0]}^{[\lambda]}(\pr P)\sum_{l=1}^{\lambda_{1}}\delta_{i\pr Pl}\\
\times\prod_{j\neq l}^{\lambda_{1}}|\uparrow(\pr Pj)\rangle\prod_{j=\lambda_{1}+1}^{N}|\downarrow(\pr Pj)\rangle.
\end{multline*}
Substituting $\pr P=\pr Q\pr P_{l\lambda_{1}}$we get
\begin{multline*}
\langle\uparrow(i)|\Xi_{tS}^{(S)}\rangle=C_{SS}\sum_{\pr Q}\sum_{l=1}^{\lambda_{1}}D_{t[0]}^{[\lambda]}(\pr Q\pr P_{l\lambda_{1}})\delta_{i\pr Q\lambda_{1}}\\
\times\prod_{j=1}^{\lambda_{1}-1}|\uparrow(\pr Qj)\rangle\prod_{j=\lambda_{1}+1}^{N}|\downarrow(\pr Qj)\rangle.
\end{multline*}
The permutation $\pr P_{l\lambda_{1}}$ permute symbols in the first
row of the Young tableau $[0]$. Therefore, $D_{t[0]}^{[\lambda]}(\pr Q\pr P_{l\lambda_{1}})=D_{t[0]}^{[\lambda]}(\pr Q)$
{[}see \refeq{DPpPpp}{]} , the summand in the equation above is independent
of $l$, and the projection can be expressed as
\begin{multline*}
\langle\uparrow(i)|\Xi_{tS}^{(S)}\rangle=\lambda_{1}C_{SS}\sum_{\pr Q}D_{t[0]}^{[\lambda]}(\pr Q)\delta_{i\pr Q\lambda_{1}}\\
\times\prod_{j=1}^{\lambda_{1}-1}|\uparrow(\pr Qj)\rangle\prod_{j=\lambda_{1}+1}^{N}|\downarrow(\pr Qj)\rangle.
\end{multline*}
The projection involved into matrix elements of $\hat{U}_{-1}$ is
evaluated in the same way,
\begin{multline*}
\langle\downarrow(i)|\Xi_{tS}^{(S)}\rangle=\lambda_{2}C_{SS}\sum_{\pr Q}D_{t[0]}^{[\lambda]}(\pr Q)\delta_{i\pr Q(\lambda_{1}+1)}\\
\times\prod_{j=1}^{\lambda_{1}}|\uparrow(\pr Qj)\rangle\prod_{j=\lambda_{1}+2}^{N}|\downarrow(\pr Qj)\rangle.
\end{multline*}

In the spin matrix elements of $\hat{U}_{\uparrow}$,
\begin{multline*}
\langle\Xi_{t'S'}^{(S')}|\uparrow(i)\rangle\langle\uparrow(i)|\Xi_{tS}^{(S)}\rangle=\delta_{SS'}\left[\lambda_{1}C_{SS}\right]^{2}\sum_{\pr Q}D_{t[0]}^{[\lambda]}(\pr Q)\delta_{i\pr Q\lambda_{1}}\\
\times\sum_{\pr R}D_{t'[0]}^{[\lambda]}(\pr R)\delta_{i\pr R\lambda_{1}}\sum_{\pr P',\pr P''}\delta_{\pr R,\pr Q\pr P'\pr P''}
\end{multline*}
the permutations $\pr R$ and $\pr Q$ can be different by permutations
$\pr P'$ of the first $\lambda_{1}-1$ symbols and $\pr P''$ of
the last $\lambda_{2}$ ones. As the permutations $\pr P'$ and $\pr P''$
do not permute symbols between rows in the Young tableau $[0]$, we
have $D_{t'[0]}^{[\lambda]}(\pr R)=D_{t'[0]}^{[\lambda]}(\pr Q)$
{[}see \refeq{DPpPpp}{]}. Since the numbers of permutations $\pr P'$
and $\pr P''$ are $(\lambda_{1}-1)!$ and $\lambda_{2}!$, respectively,
the spin matrix elements take the form,

\begin{multline*}
\langle\Xi_{t'S}^{(S)}|\uparrow(i)\rangle\langle\uparrow(i)|\Xi_{tS}^{(S)}\rangle=(\lambda_{1}-1)!\lambda_{2}!\lambda_{1}^{2}C_{SS}^{2}\\
\times\sum_{\pr Q}D_{t[0]}^{[\lambda]}(\pr Q)D_{t'[0]}^{[\lambda]}(\pr Q)\delta_{i\pr Q\lambda_{1}}.
\end{multline*}
Let us substitute this equation and \refneq{UuptilPhi} into \refneq{UuptilPsi},
perform the summation over $t$ and $t'$ using \refeq{RepProd},
and substitute $\pr P=\pr Q^{-1}\pr R^{-1}$, $j=\pr Ri$. Then the
Kronecker symbol leads to $\pr Pj=\pr Q^{-1}i=\lambda_{1}$, and we
get

\begin{multline}
\langle\tilde{\Psi}_{r'\{n'\}S}^{(S)}|\hat{U}_{\uparrow}|\tilde{\Psi}_{r\{n\}S}^{(S)}\rangle=\lambda_{1}!\lambda_{2}!\lambda_{1}C_{SS}^{2}\sum_{\pr P}D_{[0]r'}^{[\lambda]}(\pr P)D_{[0]r}^{[\lambda]}(\pr P)\\
\times\sum_{j=1}^{N}\delta_{\lambda_{1}\pr Pj}\langle n'_{j}|U|n_{j}\rangle\prod_{j'\neq j}\delta_{n'_{j'},n_{j'}}.\label{UutilPsiSum}
\end{multline}
The matrix element
\begin{multline}
\langle\tilde{\Psi}_{r'\{n'\}S-1}^{(S-1)}|\hat{U}_{-1}|\tilde{\Psi}_{r\{n\}S}^{(S)}\rangle=\frac{1}{\sqrt{2}}\lambda_{1}!\lambda_{2}!(\lambda_{2}+1)C_{SS}C_{S-1S-1}\\
\times\sum_{\pr P}D_{[0]r'}^{[\lambda']}(\pr P)D_{[0]r}^{[\lambda]}(\pr P)\sum_{j=1}^{N}\delta_{\lambda_{1}\pr Pj}\langle n'_{j}|U|n_{j}\rangle\prod_{j'\neq j}\delta_{n'_{j'},n_{j'}},\label{UmtilPsiSum}
\end{multline}
where $\lambda'=[\lambda_{1}-1,\lambda_{2}+1]$, is calculated in
a similar way.

The explicit expressions \refneq{UutilPsiSum} and \refneq{UmtilPsiSum}
are rather complicated as they include Young orthogonal matrices and
summation over all permutations. The next section provides expressions
for sums of the matrix elements and their squared moduli, which are
much simpler.

\subsection{Sum rules}

The sum of diagonal in total spin $S$ and $r$ matrix elements can
be written out as\begin{subequations}\label{SumU}
\begin{multline}
\sum_{r}\langle\tilde{\Psi}_{r'\{n'\}S}^{(S)}|\hat{U}_{a}|\tilde{\Psi}_{r\{n\}S}^{(S)}\rangle=Y^{(S)}[\hat{U}_{a}]\frac{f_{S}}{N}\\
\times\sum_{j=1}^{N}\langle n'_{j}|U|n_{j}\rangle\prod_{j'\neq j}\delta_{n'_{j'},n_{j'}}
\end{multline}
The universal factors $Y^{(S)}$ are independent of the matrix elements
$\langle n'_{j}|U|n_{j}\rangle$. For $\hat{U}_{\uparrow}$, the factor
$Y^{(S)}[\hat{U}_{\uparrow}]$ can be derived from \refneq{UutilPsiSum}
using the equalities $\sum_{r}D_{[0]r}^{[\lambda]}(\pr P)D_{[0]r}^{[\lambda]}(\pr P)=D_{[0][0]}^{[\lambda]}(\pr E)=1$
{[}obtained with \refneq{RepProd} and \refneq{InvOrthMat} {]} and
$\sum_{\pr P}\delta_{\lambda_{1}\pr Pj}=(N-1)!$, as
\begin{equation}
Y^{(S)}[\hat{U}_{\uparrow}]=\frac{N}{2}+S.\label{UupSSnpn}
\end{equation}
It is equal to the number of the spin-up atoms. For the spherical
vector component $\hat{U}_{0}$, the factor $Y^{(S)}[\hat{U}_{0}]$
is obtained using \refeq{Uupdown},
\begin{equation}
Y^{(S)}[\hat{U}_{0}]=S.\label{SumU0tilPsi}
\end{equation}
Equation \refneq{UtilPsi} leads to
\begin{equation}
Y^{(S)}[\hat{U}]=N.
\end{equation}
 \end{subequations}

The sum of squared moduli of the matrix elements \refneq{UutilPsiSum}
and \refneq{UmtilPsiSum} can be expressed, using Eqs. \refneq{flambda}
and \refneq{CSSz}, as\begin{widetext}
\begin{equation}
\sum_{r,r'}|\langle\tilde{\Psi}_{r'\{n'\}S}^{(S)}|\hat{U}_{\uparrow}|\tilde{\Psi}_{r\{n\}S}^{(S)}\rangle|^{2}=\left(\frac{\lambda_{1}f_{S}}{N!}\right)^{2}\sum_{jj'}\varSigma_{jj'}^{(S,S)}\langle n'_{j}|U|n_{j}\rangle\langle n{}_{j'}|U|n'_{j'}\rangle\prod_{j''\neq j}\delta_{n'_{j''},n_{j''}}\prod_{j'''\neq j'}\delta_{n'_{j'''},n_{j'''}}\label{SumUup2}
\end{equation}
\begin{equation}
\sum_{r,r'}|\langle\tilde{\Psi}_{r'\{n'\}S-1}^{(S-1)}|\hat{U}_{-1}|\tilde{\Psi}_{r\{n\}S}^{(S)}\rangle|^{2}=\frac{\lambda_{1}(\lambda_{2}+1)f_{S}f_{S-1}}{2(N!)^{2}}\sum_{jj'}\varSigma_{jj'}^{(S-1,S)}\langle n'_{j}|U|n_{j}\rangle\langle n{}_{j'}|U|n'_{j'}\rangle\prod_{j''\neq j}\delta_{n'_{j''},n_{j''}}\prod_{j'''\neq j'}\delta_{n'_{j'''},n_{j'''}},\label{SumUm2}
\end{equation}
 where
\begin{equation}
\varSigma_{jj'}^{(S',S)}=\sum_{r,r'}\sum_{\pr P}D_{[0]r'}^{[\lambda']}(\pr P)D_{[0]r}^{[\lambda]}(\pr P)\delta_{\lambda_{1}\pr Pj}\sum_{\pr Q}D_{[0]r'}^{[\lambda']}(\pr Q)D_{[0]r}^{[\lambda]}(\pr Q)\delta_{\lambda_{1}\pr{\pr Q}j'}.\label{SigmaSSp}
\end{equation}
\end{widetext}These sums are calculated in Appendix \ref{AppSigmaSSp}.
It is shown that 
\begin{align}
\varSigma_{jj}^{(S,S)} & =\frac{N!(N-1)!}{f_{S}\lambda_{1}^{2}}\left[\lambda_{1}-\frac{\lambda_{2}}{\lambda_{1}-\lambda_{2}+2}\right]\label{SigmaSS}\\
\varSigma_{jj}^{(S-1,S)} & =\frac{N!(N-1)!}{f_{S}\lambda_{1}}\label{SigmaSm1S}
\end{align}
 are independent of $j$, and

\begin{equation}
\varSigma_{jj'}^{(S',S)}=\frac{N!(N-2)!}{f_{S}}\delta_{SS'}-\frac{1}{N-1}\varSigma_{jj}^{(S',S)}\label{Sigmajjp}
\end{equation}
 for any $j'\neq j$.

If the sets of spatial quantum numbers $\{n\}$ and $\{n'\}$ are
different, the product of Kronecker symbols in \refneq{SumUup2} and
\refneq{SumUm2} does not vanish only if $j=j'$. Then the sum of
squared moduli of the matrix elements can be written out as\begin{subequations}\label{SumU2nnp}
\begin{multline}
\sum_{r,r'}|\langle\tilde{\Psi}_{r'\{n'\}S}^{(S')}|\hat{U}_{a}|\tilde{\Psi}_{r\{n\}S}^{(S)}\rangle|^{2}=Y^{(S,1)}[\hat{U}_{a},\hat{U}_{a}]\frac{f_{S'}}{N}\\
\times\sum_{j=1}^{N}|\langle n'_{j}|U|n_{j}\rangle|^{2}\prod_{j'\neq j}\delta_{n'_{j'},n_{j'}},
\end{multline}
where $S'\leq S$ and the difference $S-S'$ is unambiguously determined
by the operator $\hat{U}_{a}$. Each term in the sum here changes
one spatial quantum number, conserving other ones. If $U(\mathbf{r})=\mathrm{const}$,
the sums vanish since $\langle\varphi_{n'}|U|\varphi_{n}\rangle=U\langle\varphi_{n'}|\varphi_{n}\rangle=0$
for $n\neq n'$. The universal factors $Y^{(S,1)}[\hat{U}_{a},\hat{U}_{a}]$,
which are independent of the matrix elements $\langle n'_{j}|U|n_{j}\rangle$,
are expressed in terms of $\varSigma_{jj}^{(S',S)}$. Then Eqs. \refneq{SigmaSS}
and \refneq{SigmaSm1S} lead to
\begin{align}
Y^{(S,1)}[\hat{U}_{\uparrow},\hat{U}_{\uparrow}] & =\frac{N}{2}+S-\frac{N-2S}{4(S+1)}\label{Uup2SSnpn}\\
Y^{(S,1)}[\hat{U}_{-1},\hat{U}_{-1}] & =\frac{N-2S+2}{4},\label{Uup2Sm1Snpn}
\end{align}
and \refeq{UtilPsi} gives 
\[
Y^{(S,1)}[\hat{U},\hat{U}]=N.
\]
 The factor $Y^{(S,1)}[\hat{U}_{0},\hat{U}_{0}]$ for the spherical
vector component $\hat{U}_{0}$ is obtained using \refneq{Uupdown}.
Since the matrix elements of $\hat{U}$ are diagonal in $r$ {[}see
\refeq{UtilPsi}{]}, one gets
\begin{equation}
Y^{(S,1)}[\hat{U}_{0},\hat{U}_{0}]=\frac{S(N+2)}{4(S+1)}.
\end{equation}
\end{subequations}

For transitions conserving the spatial quantum numbers, $\{n'\}=\{n\}$
and the Kronecker symbols in \refneq{SumUup2} and \refneq{SumUm2}
are equal to one for any $j$ and $j'$. Then sums of squared moduli
of the matrix elements can be represented as\begin{subequations}\label{SumU2nn}
\begin{multline}
\sum_{r,r'}|\langle\tilde{\Psi}_{r'\{n\}S'}^{(S')}|\hat{U}_{a}|\tilde{\Psi}_{r\{n\}S}^{(S)}\rangle|^{2}=f_{S'}\Bigl[Y_{0}^{(S,0)}[\hat{U}_{a},\hat{U}_{a}]\langle U\rangle^{2}\\
+Y_{1}^{(S,0)}[\hat{U}_{a},\hat{U}_{a}]\langle\Delta U\rangle^{2}\Bigr]
\end{multline}
 where
\[
\langle U\rangle=\frac{1}{N}\sum_{j=1}^{N}\langle n_{j}|U|n_{j}\rangle
\]
is the average matrix element and
\[
\langle\Delta U\rangle=\left[\frac{1}{N}\sum_{j=1}^{N}\left(\langle n_{j}|U|n_{j}\rangle-\langle U\rangle\right)^{2}\right]^{1/2}
\]
is the average deviation of the matrix elements of $U(\mathbf{r})$.
The universal factors $Y_{0}^{(S,0)}[\hat{U}_{a},\hat{U}_{a}]$ and
$Y_{1}^{(S,0)}[\hat{U}_{a},\hat{U}_{a}]$ are independent of the matrix
elements $\langle n_{j}|U|n_{j}\rangle$. Equations \refneq{SigmaSS},
\refneq{SigmaSm1S}, and \refneq{Sigmajjp} lead to
\begin{equation}
Y_{0}^{(S,0)}[\hat{U}_{a},\hat{U}_{a}]=\left(Y^{(S)}[\hat{U}_{a}]\right)^{2}
\end{equation}
(where defined $Y^{(S)}[\hat{U}_{-1}]=0$, in addition to \refeq{SumU}),
and 
\begin{align}
Y_{1}^{(S,0)}[\hat{U}_{-1},\hat{U}_{-1}] & =\frac{N(N-2S+2)}{4(N-1)}\\
Y_{1}^{(S,0)}[\hat{U}_{\uparrow},\hat{U}_{\uparrow}] & =Y_{1}^{(S,0)}[\hat{U}_{0},\hat{U}_{0}]\nonumber \\
 & =\frac{S(N-2S)(N+2S+2)}{4(S+1)(N-1)}.
\end{align}
\end{subequations}If $U(\mathbf{r})=\mathrm{const}$, $\Delta U=0$,
and, therefore, $\sum_{r,r'}|\langle\tilde{\Psi}_{r'\{n\}S-1}^{(S-1)}|\hat{U}_{-1}|\tilde{\Psi}_{r\{n\}S}^{(S)}\rangle|^{2}=0$.
Indeed, in this case, the spatial matrix elements \refneq{UuptilPhi}
are equal to zero due to the orthogonality of the spatial wavefunctions
with different spins.

Thus, sums of matrix elements and their squared moduli are expressed
in terms of universal factors, which are independent of the spatial
orbitals and details of the external fields, and sums of one-body
matrix elements (or their squared moduli), which are independent of
many-body spins. The sum rules, combined with the spin-projection
dependence \refneq{UvectWE}, provide information on each matrix element
for an one-body spin-dependent interaction with an external field.

\section{Sum rules for two-body spin-independent interactions\label{secTwoBody}}

The permutation-invariant interaction between particles is given by
\begin{equation}
\hat{V}=\sum_{j\neq j'}V(\mathbf{r}_{j}-\mathbf{r}_{j'}).\label{VindepS}
\end{equation}
Without loss of generality, we can restrict consideration to even
potential functions, $V(\mathbf{r})=V(\mathbf{-r})$, since their
odd parts are canceled. Matrix elements of this interaction can be
evaluated for general spin wavefunctions. Due to the orthogonality
of the spin wavefunctions \refneq{Xiorth}, the matrix elements are
diagonal in spin quantum numbers and can be reduced to the matrix
elements between spatial wavefunctions,\begin{widetext}
\begin{equation}
\langle\tilde{\Psi}_{r'\{n'\}l'}^{(S')}|\hat{V}|\tilde{\Psi}_{r\{n\}l}^{(S)}\rangle=\delta_{SS'}\delta_{ll'}\frac{2}{f_{S}}\sum_{t}\sum_{i<i'}\langle\tilde{\Phi}_{tr'\{n'\}}^{(S)}|V(\mathbf{r}_{i}-\mathbf{r}_{i'})|\tilde{\Phi}_{tr\{n\}}^{(S)}\rangle.\label{VtilPsitilPhi}
\end{equation}
(this reduction is used in spin-free quantum chemistry \cite{kaplan,pauncz_symmetric}).
Then, using \refneq{tilPhi}, \refneq{VindepS}, and the property
\refneq{InvOrthMat} of the Young orthogonal matrices, the spatial
matrix elements can be expressed as
\begin{multline}
\langle\tilde{\Phi}_{tr'\{n'\}}^{(S)}|V(\mathbf{r}_{i}-\mathbf{r}_{i'})|\tilde{\Phi}_{tr\{n\}}^{(S)}\rangle=\frac{f_{S}}{N!}\sum_{\pr R,\pr Q}\mathrm{sgn}(\pr Q)D_{r't}^{[\lambda]}(\pr Q)\mathrm{sgn}(\pr R)D_{rt}^{[\lambda]}(\pr R)\\
\times\int d^{D}r_{i}d^{D}r_{i'}\varphi_{n'_{\pr Qi}}^{*}(\mathbf{r}_{i})\varphi_{n'_{\pr Qi'}}^{*}(\mathbf{r}_{i'})V(\mathbf{r}_{i}-\mathbf{r}_{i'})\varphi_{n_{\pr Ri}}(\mathbf{r}_{i})\varphi_{n_{\pr Ri'}}(\mathbf{r}_{i'})\prod_{i'\neq i''\neq i}\delta_{n'_{\pr Qi''},n_{\pr Ri''}}.\label{VtilPhi}
\end{multline}
 The Kronecker $\delta$-symbols appear here due to the orthogonality
of the spatial orbitals $\varphi_{n}$ and the absence of equal quantum
numbers in each of the sets $\{n\}$ and $\{n'\}$. Due to the $\delta$-symbols,
all but two spatial quantum numbers remain unchanging. Supposing that
the unchanged $n_{i''}$ are in the same positions in the sets $\{n\}$
and $\{n'\}$, one can see that the Kronecker symbols allow only $\pr Q=\pr R$
or $\pr Q=\pr R\pr P_{ii'}$. Then substitution of \refneq{VtilPhi}
into \refneq{VtilPsitilPhi}, using \refneq{RepProd} and \refneq{InvOrthMat},
leads to
\begin{multline}
\langle\tilde{\Psi}_{r'\{n'\}l'}^{(S')}|\hat{V}|\tilde{\Psi}_{r\{n\}l}^{(S)}\rangle=2\delta_{SS'}\delta_{ll'}\frac{1}{N!}\sum_{\pr R}\sum_{i<i'}\prod_{\pr Ri'\neq j''\neq\pr Ri}\delta_{n'_{j''},n_{j''}}\\
\times\left[\delta_{r'r}\langle n'_{\pr Ri}n'_{\pr Ri'}|V|n_{\pr Ri}n_{\pr Ri'}\rangle+\mathrm{sgn}(\pr P_{ii'})D_{r'r}^{[\lambda]}(\pr R\pr P_{ii'}\pr R^{-1})\langle n'_{\pr Ri'}n'_{\pr Ri}|V|n_{\pr Ri}n_{\pr Ri'}\rangle\right],\label{VtilPsiSumR}
\end{multline}
where $\langle n'_{1}n'_{2}|V|n_{1}n_{2}\rangle=\int d^{D}r_{1}d^{D}r_{2}\varphi_{n'_{1}}^{*}(\mathbf{r}_{1})\varphi_{n'_{2}}^{*}(\mathbf{r}_{2})V(\mathbf{r}_{1}-\mathbf{r}_{2})\varphi_{n_{1}}(\mathbf{r}_{1})\varphi_{n_{2}}(\mathbf{r}_{2})$. 

Taking into account that 
\begin{equation}
\pr P\pr P_{ii'}\pr P^{-1}=\pr P_{\pr Pi\pr Pi'}\label{TransfTransp}
\end{equation}
 (see \cite{pauncz_symmetric}) and substituting $\pr Ri=j$, one
finally gets
\begin{multline}
\langle\tilde{\Psi}_{r'\{n'\}l'}^{(S')}|\hat{V}|\tilde{\Psi}_{r\{n\}l}^{(S)}\rangle=2\delta_{SS'}\delta_{ll'}\sum_{j<j'}\prod_{j'\neq j''\neq j}\delta_{n'_{j''},n_{j''}}\left[\delta_{r'r}\langle n'_{j}n'_{j'}|V|n_{j}n_{j'}\rangle+\mathrm{sgn}(\pr P_{jj'})D_{r'r}^{[\lambda]}(\pr P_{jj'})\langle n'_{j'}n'_{j}|V|n_{j}n_{j'}\rangle\right].\label{VtilPsi}
\end{multline}
It is a special case of the matrix elements obtained by Heitler \cite{heitler1927}
and Kaplan \cite{kaplan}. 

The sum of diagonal elements of the representation matrix, the character
\[
\chi_{S}(\pr C)\equiv\sum_{r}D_{rr}^{[\lambda]}(\pr P),
\]
is the same for all permutations $\pr P$, which form the class of
conjugate elements $\pr C$ \cite{hamermesh,elliott,kaplan,pauncz_symmetric}.
Table \ref{Tab_char} presents the characters for the classes appearing
here. (Supplemental material for \cite{yurovsky2014} contains a code
based on the explicit expressions \cite{lassalle2008} for the characters.)
The conjugated classes of the symmetric group $\pr S_{N}$ are characterized
by the cyclic structure of the permutations. All permutations in the
class $\pr C=\{N^{\nu_{N}}\ldots2^{\nu_{2}}\}$ have $\nu_{l}$ cycles
of length $l$. This class notation omits $l^{\nu_{l}}$ if $\nu_{l}=0$
and the number of cycles of the length one, i.e. the number of symbols
which are not affected by the permutations in the class. This number
is determined by the condition $\sum_{l=1}^{N}l\nu_{l}=N$. Permutations
of two symbols form the class $\{2\}$. This leads to the sum of diagonal
in $r$ matrix elements\begin{subequations}\label{SumV} 
\begin{equation}
\sum_{r}\langle\tilde{\Psi}_{r\{n'\}l}^{(S)}|\hat{V}|\tilde{\Psi}_{r\{n\}l}^{(S)}\rangle=2\sum_{j<j'}\left[f_{S}\langle n'_{j}n'_{j'}|V|n_{j}n_{j'}\rangle\pm\chi_{S}(\{2\})\langle n'_{j'}n'_{j}|V|n_{j}n_{j'}\rangle\right]\prod_{j'\neq j''\neq j}\delta_{n'_{j''},n_{j''}},\label{SumVnnp}
\end{equation}
where the sign $+$ or $-$ is taken for bosons or fermions, respectively.
Similar expressions have been obtained for the total energy \cite{heitler1927}
and arbitrary observables \cite{yurovsky2014}. If $\{n'\}=\{n\}$,
the Kronecker symbols are equal to one for any $j$ and $j'$ and
the sum can be transformed to the form
\begin{equation}
\sum_{r}\langle\tilde{\Psi}_{r\{n\}l}^{(S)}|\hat{V}|\tilde{\Psi}_{r\{n\}l}^{(S)}\rangle=N(N-1)f_{S}\left(\langle V\rangle_{\mathrm{dir}}\pm\frac{\chi_{S}(\{2\})}{f_{S}}\langle V\rangle_{\mathrm{ex}}\right).\label{sumVtilPsi}
\end{equation}
\end{subequations}Here and above, the dependence on many-body states
is given by universal functions $f_{S}$ and $\chi_{S}(\{2\})$, which
are independent of the matrix elements $\langle n'_{1}n'_{2}|V|n_{1}n_{2}\rangle$,
while the average matrix elements 
\begin{equation}
\langle V\rangle_{\mathrm{dir}}=\frac{2}{N(N-1)}\sum_{j<j'}\langle n_{j}n_{j'}|V|n_{j}n_{j'}\rangle,\quad\langle V\rangle_{\mathrm{ex}}=\frac{2}{N(N-1)}\sum_{j<j'}\langle n_{j'}n_{j}|V|n_{j}n_{j'}\rangle\label{Vdirex}
\end{equation}
of the direct and exchange interactions, respectively, are independent
of the many-body states.

\begin{table*}
\protect\caption{Characters $\chi_{S}(\protect\pr C)$ of the classes $\protect\pr C$
of conjugate elements of the symmetric group $\protect\pr S_{N}$
of permutations of $N$ symbols in the irreducible representations,
corresponding to the spin $S$. The characters are calculated with
the Frobenius formula\cite{pauncz_symmetric,murnaghan} and scaled
to the representation dimension $f_{S}$.}

\begin{centering}
\begin{tabular}{|c|c|}
\hline 
$\pr C$  & $\chi_{S}(\pr C)/f_{S}$\tabularnewline
\hline 
$\{2\}$  & $\frac{4S^{2}+N^{2}+4S-4N}{2N(N-1)}$\tabularnewline
\hline 
$\{3\}$  & $\frac{12S^{2}+N^{2}+12S-10N}{4N(N-1)}$\tabularnewline
\hline 
$\{4\}$  & $\frac{N^{4}-24N^{3}+4N^{2}(6S^{2}+6S+29)-16N(10S^{2}+10S+9)+16S(S+1)(S^{2}+S+12)}{8N(N-1)(N-2)(N-3)}$\tabularnewline
\hline 
$\{2^{2}\}$  & $\frac{N^{4}-12N^{3}+8N^{2}(S^{2}+S+7)+8N(10S^{2}+10S+9)+16S(S+1)(S^{2}+S+6)}{4N(N-1)(N-2)(N-3)}$\tabularnewline
\hline 
\end{tabular}
\par\end{centering}

\label{Tab_char} 
\end{table*}

Calculating the sum of squared moduli of the matrix elements \refneq{VtilPsi},
one can see that if the sets of spatial quantum numbers $\{n\}$ and
$\{n'\}$ are different by two elements, the product of Kronecker
symbols in the product of the matrix elements does not vanish only
if the pair $j$, $j'$ is the same in both matrix elements. Then
the sum can be expressed as\begin{subequations}\label{SumV2}
\begin{multline}
\sum_{r,r'}|\langle\tilde{\Psi}_{r'\{n'\}l}^{(S)}|\hat{V}|\tilde{\Psi}_{r\{n\}l}^{(S)}\rangle|^{2}=4f_{S}\sum_{j<j'}\prod_{j'\neq j''\neq j}\delta_{n'_{j''},n_{j''}}\biggl[|\langle n'_{j}n'_{j'}|V|n_{j}n_{j'}\rangle|^{2}+|\langle n'_{j'}n'_{j}|V|n_{j}n_{j'}\rangle|^{2}\\
\pm2\frac{\chi_{S}(\{2\})}{f_{S}}\mathrm{Re}\left(\langle n'_{j}n'_{j'}|V|n_{j}n_{j'}\rangle\langle n'_{j'}n'_{j}|V|n_{j}n_{j'}\rangle^{*}\right)\biggr].\label{SumV2nnp}
\end{multline}
Here the equality $\sum_{rr'}D_{r'r}^{[\lambda]}(\pr P_{jj'})D_{r'r}^{[\lambda]}(\pr P_{jj'})=\sum_{r}D_{rr}^{[\lambda]}(\pr E)=f_{S}$
was used. Each term in the sum above changes two of the spatial quantum
numbers, conserving other ones. The case of a single changed quantum
number will be considered elsewhere.

For transitions conserving the spatial quantum numbers, $\{n'\}=\{n\}$
and the Kronecker symbols in \refneq{VtilPsi} are equal to one for
any $j$ and $j'$. Then
\begin{multline*}
\sum_{r,r'}|\langle\tilde{\Psi}_{r'\{n\}l}^{(S)}|\hat{V}|\tilde{\Psi}_{r\{n\}l}^{(S)}\rangle|^{2}=\Bigl[f_{S}N^{2}(N-1)^{2}\langle V\rangle_{\mathrm{dir}}^{2}\pm2\chi_{S}(\{2\})N^{2}(N-1)^{2}\langle V\rangle_{\mathrm{dir}}\langle V\rangle_{\mathrm{ex}}\\
+\sum_{j_{1}\neq j_{1}'}\sum_{j_{2}\neq j_{2}'}\sum_{r}D_{rr}^{[\lambda]}(\pr P_{j_{1}j'_{1}}\pr P_{j_{2}j'_{2}})\langle n_{j_{1}'}n_{j_{1}}|V|n_{j_{1}}n_{j_{1}'}\rangle\langle n_{j_{2}'}n_{j_{2}}|V|n_{j_{2}}n_{j_{2}'}\rangle\Bigr]
\end{multline*}
The trace of the Young matrix can be transformed in the following
way (since $j_{1}\neq j'_{1}$ and $j_{2}\neq j'_{2}$)
\begin{multline*}
\sum_{r}D_{rr}^{[\lambda]}(\pr P_{j_{1}j'_{1}}\pr P_{j_{2}j'_{2}})=\chi_{S}(\{2^{2}\})+(\delta_{j_{1}j_{2}}+\delta_{j_{1}j'_{2}}+\delta_{j'_{1}j_{2}}+\delta_{j'_{1}j'_{2}})(\chi_{S}(\{3\})-\chi_{S}(\{2^{2}\}))\\
+(\delta_{j_{1}j_{2}}\delta_{j'_{1}j'_{2}}+\delta_{j_{1}j'_{2}}\delta_{j'_{1}j_{2}})(f_{S}-2\chi_{S}(\{3\})+\chi_{S}(\{2^{2}\})),
\end{multline*}
 since $\pr P_{j_{1}j'_{1}}\pr P_{j_{1}j'_{2}}\in\{3\}$ for $j'_{1}\neq j'_{2}$,
$\pr P_{j_{1}j'_{1}}\pr P_{j_{1}j'_{1}}=\pr E$, and $\chi_{S}(\pr E)=f_{S}$.
Here and in what follows, $\chi_{S}(\{3\})$ and $\chi_{S}(\{2^{2}\})$
have to be equated to zero at $N<3$ and $N<4$, respectively, when
the corresponding permutations do not exist. Using the identity $2f_{S}+4(N-2)\chi_{S}(\{3\})+(N-2)(N-3)\chi_{S}(\{2^{2}\})=N(N-1)\chi_{S}^{2}(\{2\})/f_{S}$
(it can be directly proved with the characters in Table \ref{Tab_char}),
the sum of squared moduli of the matrix elements can be represented
as
\begin{equation}
\sum_{r,r'}|\langle\tilde{\Psi}_{r'\{n\}l}^{(S)}|\hat{V}|\tilde{\Psi}_{r\{n\}l}^{(S)}\rangle|^{2}=f_{S}\left(Y_{1}^{(S,0)}[\hat{V},\hat{V}]\langle\Delta_{1}V\rangle^{2}+Y_{2}^{(S,0)}[\hat{V},\hat{V}]\langle\Delta_{2}V\rangle^{2}\right)+\frac{1}{f_{S}}\left(\sum_{r}\langle\tilde{\Psi}_{r\{n\}l}^{(S)}|\hat{V}|\tilde{\Psi}_{r\{n\}l}^{(S)}\rangle\right)^{2}\label{sumVtilPsi2}
\end{equation}
with the universal factors
\begin{equation}
Y_{1}^{(S,0)}[\hat{V},\hat{V}]=4N(N-1)^{2}\frac{\chi_{S}(\{3\})-\chi_{S}(\{2^{2}\})}{f_{S}},\quad Y_{2}^{(S,0)}[\hat{V},\hat{V}]=2N(N-1)\left(1-\frac{2\chi_{S}(\{3\})-\chi_{S}(\{2^{2}\})}{f_{S}}\right).\label{YSS2}
\end{equation}
\end{subequations}\end{widetext}Here
\begin{equation}
\begin{split}\langle\Delta_{1}V\rangle^{2}= & \frac{1}{N}\sum_{j=1}^{N}\left(\frac{1}{N-1}\sum_{j'\neq j}\langle n_{j'}n_{j}|V|n_{j}n_{j'}\rangle-\langle V\rangle_{\mathrm{ex}}\right)^{2}\\
\langle\Delta_{2}V\rangle^{2}= & \frac{2}{N(N-1)}\sum_{j<j'}(\langle n_{j'}n_{j}|V|n_{j}n_{j'}\rangle-\langle V\rangle_{\mathrm{ex}})^{2}
\end{split}
\label{DeltaV12}
\end{equation}
measure the average deviation of the exchange matrix elements.

Thus, sums of matrix elements and their squared moduli are expressed
in terms of universal factors, which are independent of the spatial
orbitals and interaction potentials, and sums of two-body matrix elements
(or their squared moduli), which are independent of the many-body
spins. The universal factors are expressed in terms of characters
of irreducible representations of the symmetric group. The characters
are functions of the total spin and the number of particles.

\section{Multiplet energies for weakly-interacting gases\label{sec_energies}}

As an example of applications of the sum rules, consider splitting
of degenerate energy levels due to weak two-body spin-independent
interactions. The Hamiltonian of the system is a sum of one-body Hamiltonians
$\hat{H}_{0}(j)$ of non-interacting particles and two-body interactions
\refneq{VindepS}, 
\begin{equation}
\hat{H}_{\mathrm{spat}}=\sum_{j=1}^{N}\hat{H}_{0}(j)+\hat{V}\label{HspatV}
\end{equation}
The interactions split energies of the degenerate states \refneq{tilPsiSrnl}.
In the zero-order of the degenerate perturbation theory \cite{landau},
the eigenenergies $E_{Sn}$ (counted from the multiplet-independent
energy of non-interacting particles $\sum_{j=1}^{N}\varepsilon_{n_{j}}$)
are determined by the secular equation
\begin{equation}
\sum_{r'}V_{rr'}^{(S)}A_{nr'}^{(S)}=E_{Sn}A_{nr}^{(S)},\label{secular}
\end{equation}
where $A_{nr}^{(S)}$ are the expansion coefficients of the wavefunction
\refneq{PsiSnSz} in terms of the wavefunctions of non-interacting
particles \refneq{tilPsiSrnSz},
\begin{equation}
\Psi_{nS_{z}}^{(S)}=\sum_{r}A_{nr}^{(S)}\tilde{\Psi}_{r\{n\}S_{z}}^{(S)}\label{PsiInt}
\end{equation}
and the matrix elements of the spin-independent two-body interaction
\refneq{VtilPsi} 
\[
V_{rr'}^{(S)}=\langle\tilde{\Psi}_{r'\{n\}S_{z}}^{(S)}|\hat{V}|\tilde{\Psi}_{r\{n\}S_{z}}^{(S)}\rangle
\]
do not couple states with different spins. 

Consider at first the case when the matrix elements $V_{\mathrm{dir}}=\langle n_{1}n_{2}|\hat{V}|n_{1}n_{2}\rangle$
and $V_{\mathrm{ex}}=\langle n_{1}n_{2}|\hat{V}|n_{2}n_{1}\rangle$
are independent of the spatial quantum numbers. E.g., this can take
place in the case of zero-range interactions $V(\mathbf{r})=V\delta(\mathbf{r})$,
if the spatial orbitals have a form of plane waves. In this case,
the summation over $\pr R$ in the matrix element \refneq{VtilPsiSumR}
for $\{n\}=\{n'\}$ can be performed using Eqs. \refneq{RepProd},
\refneq{InvOrthMat}, and the orthogonality relation \refneq{OrthRel}
in the following way \cite{elliott} 
\begin{multline*}
\sum_{\pr R}D_{r'r}^{[\lambda]}(\pr R\pr P_{ii'}\pr R^{-1})=\sum_{t,t'}D_{t't}^{[\lambda]}(\pr P_{ii'})\sum_{\pr R}D_{r't'}^{[\lambda]}(\pr R)D_{rt}^{[\lambda]}(\pr R)\\
=\frac{N!}{f_{S}}\delta_{r'r}\chi_{S}(\{2\}).
\end{multline*}
Then the matrix elements become diagonal in $r$,
\[
V_{rr'}^{(S)}=\delta_{rr'}N(N-1)\left(V_{\mathrm{dir}}\pm\frac{\chi_{S}(\{2\})}{f_{S}}V_{\mathrm{ex}}\right),
\]
where the character $\chi_{S}(\{2\})$ is given in Tab. \ref{Tab_char}
and the sign $+$ or $-$ is taken for bosons or fermions, respectively.
The secular equation \refneq{secular} is then satisfied by the eigenvectors
$A_{nr}^{(S)}=\delta_{nr}$ and eigenvalues $E_{Sn}=V_{rr}^{(S)}$.
Then all eigenstates with the given spin remain degenerate in energy.
However, states with different total spins have different energies.
\begin{figure}
\includegraphics[width=3.4in]{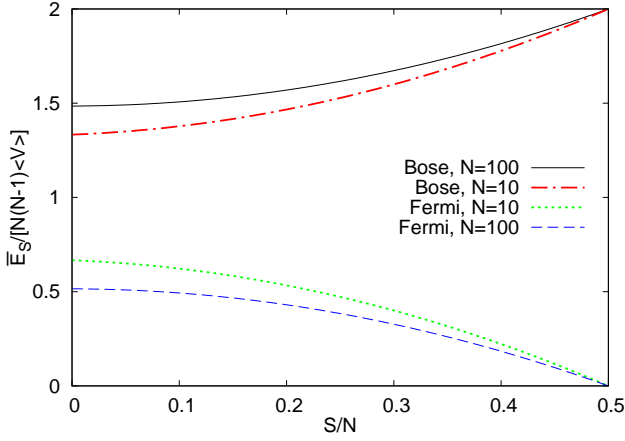}

\protect\caption{(Color online) Scaled average multiplet energies $\bar{E}_{S}$ as
functions of the multiplet spin $S$ for $N$ bosons or fermions.
The energies are calculated with \protect\refeq{barES} and scaled
characters from Tab.\ \ref{Tab_char}, assuming $\langle V\rangle_{\mathrm{dir}}=\langle V\rangle_{\mathrm{ex}}=\langle V\rangle$.\label{FigEaver}}
\end{figure}

In the general case, when the matrix elements of $\hat{V}$ depend
on the spatial quantum numbers, the energies $E_{Sn}$ can not be
expressed in a simple form. However, using the equivalence of the
sum of matrix eigenvalues to its trace and the sum of matrix elements
\refneq{sumVtilPsi}, the average multiplet energy can be expressed
as 
\begin{multline}
\bar{E}_{S}\equiv\frac{1}{f_{S}}\sum_{n}E_{Sn}=\frac{1}{f_{S}}\sum_{r}V_{rr}\\
=N(N-1)\left(\langle V\rangle_{\mathrm{dir}}\pm\frac{\chi_{S}(\{2\})}{f_{S}}\langle V\rangle_{\mathrm{ex}}\right),\label{barES}
\end{multline}
where the average interactions $\langle V\rangle_{\mathrm{dir}}$
and $\langle V\rangle_{\mathrm{ex}}$ are defined by \refneq{Vdirex}.
(Here and below, the summation over $n$ means the summation over
states of interacting particles in a given spin-multiplet with a given
set $\{n\}$.) It is a particular case of the general expression obtained
by Heitler \cite{heitler1927}. The average energies are plotted in
Fig. \ref{FigEaver}.

As the interaction lifts degeneracy of states with different total
spins, transformation of the set of states with defined total spins
to the set of states with given spin projections of particles becomes
impossible. Then the former set remains the only valid set of eigenstates
of interacting particles. 

For fermions, the average multiplet energy decreases with $S$. The
Lieb-Mattis theorem \cite{lieb1962} predicts opposite dependence.
However, this theorem is formulated for the lowest-energy states with
given $S$, which can involve different sets of $\{n\}$ and have
multiple occupation of spatial orbitals. In contrast, the average
energies \refneq{barES} are obtained for the fixed set of $\{n\}$
and single occupations. 
\begin{figure}
\includegraphics[width=3.4in]{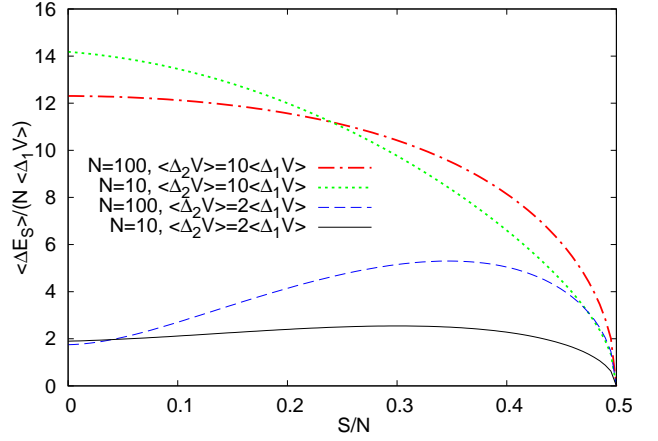}

\protect\caption{(Color online) Scaled root-mean-square energy widths of multiplets
as functions of the multiplet spin $S$ for $N$ particles, calculated
with \protect\refeq{DeltaES2}.\label{FigDeltaE}}
\end{figure}

The root-mean-square energy width of the spin-$S$ multiplet $\langle\Delta E_{S}\rangle$
is defined by 
\[
\langle\Delta E_{S}\rangle^{2}\equiv\frac{1}{f_{S}}\sum_{n}(E_{Sn}-\bar{E}_{S})^{2}=\frac{1}{f_{S}}\sum_{n}E_{Sn}^{2}-\bar{E}_{S}^{2}
\]

Due to orthogonality of the expansion coefficients, the secular equation
\refneq{secular} can be rewritten in the form $E_{Sn}\delta_{n'n}=\sum_{rr'}A_{nr}^{*}V_{rr'}A_{n'r'}$,
leading to
\[
\frac{1}{f_{S}}\sum_{n,n'}|E_{Sn}\delta_{n'n}|^{2}=\frac{1}{f_{S}}\sum_{r,r'}V_{rr'}^{*}V_{r'r}.
\]
Then \refeq{sumVtilPsi2} gives us
\begin{equation}
\langle\Delta E_{S}\rangle^{2}=Y_{1}^{(S,0)}[\hat{V},\hat{V}]\langle\Delta_{1}V\rangle^{2}+Y_{2}^{(S,0)}[\hat{V},\hat{V}]\langle\Delta_{2}V\rangle^{2},\label{DeltaES2}
\end{equation}
where the universal factors $Y_{1}^{(S,0)}[\hat{V},\hat{V}]$ and
$Y_{2}^{(S,0)}[\hat{V},\hat{V}]$ are expressed in terms of the representation
characters by \refeq{YSS2}, and the matrix element deviations $\langle\Delta_{1}V\rangle$
and $\langle\Delta_{2}V\rangle$ are defined by \refeq{DeltaV12}.
The multiplet energy widths are plotted in Fig. \ref{FigDeltaE}.
If the matrix elements of $\hat{V}$ are independent of the spatial
quantum numbers, $\langle\Delta_{1}V\rangle=\langle\Delta_{2}V\rangle=0$
and, therefore, $\langle\Delta E_{S}\rangle=0$, in agreement with
the above-mentioned degeneracy of states with given $S$ in this case.
The energy width is determined by characters, which were identified
by Dirac \cite{dirac1929} as constants of motions, corresponding
to permutation symmetry, according to generalized Noether\textquoteright s
theorem. Therefore, the energy width can be considered as a conserved
physical observable, related to this symmetry, as well as the average
multiplet energy and correlations \cite{yurovsky2014}.

Using characters from Tab.\ \ref{Tab_char}, the exact expression
can be approximated at $N\gg1$ by 
\begin{multline*}
\langle\Delta E_{S}\rangle^{2}\approx\frac{N^{2}-4S^{2}}{2N^{2}}V_{1D}^{2}[2N(4S^{2}-3N)\langle\Delta_{1}V\rangle^{2}\\
+(3N^{2}-4S^{2})\langle\Delta_{2}V\rangle^{2}].
\end{multline*}

Consider now external fields described by one-body interactions. Matrix
elements of a spin-independent field \refneq{UtilPsi} are independent
of $r$ and spin quantum numbers. Therefore, this field leads to the
same shift for all states, corresponding to the given set of spatial
quantum numbers $\{n\}$. In the first order of the perturbation theory,
this shift will be $\sum_{j=1}^{N}\langle n{}_{j}|U|n_{j}\rangle$.
Even strong spin-independent field leads to the same shift of all
states, as it can be incorporated into the Hamiltonian of non-interacting
particles. Then, the Schr\"odinger equation \refneq{SchrH0} will
contain $\hat{H}_{0}(j)+U(\mathbf{r}_{j})$. This leads to different
one-body eigenfunctions $\varphi_{n}(\mathbf{r})$ and eigenvalues
$\varepsilon_{n}$, but does not change the form of many-body wavefunctions.

Spin-dependent spatially-homogeneous interactions {[}Eqs. \refneq{Uvect}
and \refneq{Uupdown} with $U=\mathrm{const}${]} commute with the
spatial Hamiltonian of interacting particles \refneq{HspatV}. Since
the spin wavefunctions \refneq{XiStSz} are eigenfunctions of such
interactions, the eigenfunctions $\Psi_{nS_{z}}^{(S)}$ of $\hat{H}_{\mathrm{spat}}$
will be eigenfunctions of the Hamiltonian $\hat{H}_{\mathrm{spat}}+\hat{U}_{0}$.
The energy shift of the states of non-interacting particles due to
the field $\hat{U}_{0}$ is equal to the matrix element $\langle\tilde{\Psi}_{r'\{n\}S_{z}}^{(S)}|\hat{U}_{0}|\tilde{\Psi}_{r\{n\}S_{z}}^{(S)}\rangle=\delta_{r'r}S_{z}U$.
It is determined by Eqs. \refneq{Uupdown}, \refneq{UvectWE}, \refneq{UtilPsi}
and \refneq{UutilPsiSum}, taking into account that $\langle n'|U|n\rangle=U\delta_{nn'}$.
The energy shift of the states of interacting particles \refneq{PsiInt}
will be the same, as $\langle\Psi_{nS_{z}}^{(S)}|\hat{U}_{0}|\Psi_{nS_{z}}^{(S)}\rangle=S_{z}U\sum_{r}A_{nr}^{*}A_{nr}=S_{z}U$. 

The spin-independent inhomogeneous and spin-dependent homogeneous
fields, considered above, are consistent with the separation \refneq{HSpinSpat}
of the spin and spatial Hamiltonians. If the external field depends
both on spins and coordinates, this separation is violated, invalidating
the use of collective spin and spatial wavefunctions for non-interacting
particles. Nevertheless, these wavefunctions remain applicable to
interacting particles whenever the external field is weak enough,
and the energy shift can be estimated in the first order of the perturbation
theory. The average shift is calculated using orthogonality of the
coefficients $A_{nr}$, Eqs. \refneq{UvectWE} and \refneq{SumU}
in the following way, 
\begin{multline}
\frac{1}{f_{S}}\sum_{n}\langle\Psi_{nS_{z}}^{(S)}|\hat{U}_{0}|\Psi_{nS_{z}}^{(S)}\rangle\\
=\frac{1}{f_{S}}\sum_{n}A_{nr'}^{*}A_{nr}\langle\tilde{\Psi}_{r'\{n\}S_{z}}^{(S)}|\hat{U}_{0}|\tilde{\Psi}_{r\{n\}S_{z}}^{(S)}\rangle\\
=X_{S_{z}0}^{(S,S,1)}Y^{(S)}[\hat{U}_{0}]\frac{1}{N}\sum_{j=1}^{N}\langle n{}_{j}|U|n_{j}\rangle=\frac{S_{z}}{N}\sum_{j=1}^{N}\langle n{}_{j}|U|n_{j}\rangle.\label{shift}
\end{multline}

\section*{Conclusions}

The symmetric group methods allow to evaluate the matrix elements
of spin-dependent external fields \refneq{Uvect} and spin-independent
two-body interactions \refneq{VindepS} in the basis with collective
spin and spatial wavefunctions \refneq{PsiSnl}. These matrix elements
agree to the selection rules \cite{yurovsky2014}. For the matrix
elements of spin-dependent external fields, explicit dependence on
the total spin projection \refneq{UvectWE} is obtained using the
Wigner-Eckart theorem. Analytical expressions are derived for sums
of these matrix elements \refneq{SumU} and their squared moduli {[}Eqs.
\refneq{SumU2nnp} and \refneq{SumU2nn}{]} over irreducible representations
for both spin-conserving and spin-changing transitions. Dependence
on the many-body states in these sums is given by the $3j$ Wigner
symbols and the universal factors $Y^{(S)}$, $Y^{(S,1)}$, $Y_{0}^{(S,0)}$,
and $Y_{1}^{(S,0)}$. These factors are independent of details of
one-body Hamiltonians and external fields and are expressed in a rather
simple form in terms of the total spin and number of particles. For
spin-independent two-body interactions, the sums of matrix elements
\refneq{SumV} and their squared moduli \refneq{SumV2} depend on
the many-body states only through the representation characters, which
were identified by Dirac \cite{dirac1929} as constants of motions,
corresponding to permutation symmetry. The sum rules can be applied
to the evaluation of energy-level shifts \refneq{shift}, splitting
of states with different total spins \refneq{barES}, and spin-multiplet
energy widths \refneq{DeltaES2}. Other possible applications of the
sum rules include estimates of the spin-multiplet depletion rates
due to spin-dependent perturbations, as well as the population transfer
rates between spin-multiplets using the spatially-homogeneous spin-changing
and spatially-inhomogeneous spin-conserving pulses \cite{yurovsky2014}.

\appendix*

\section{Calculation of the sums \refneq{SigmaSSp}\label{AppSigmaSSp}}

Using the relations \refneq{RepProd} and \refneq{InvOrthMat} and
substitution $\pr R=\pr Q\pr P^{-1}$, the sum \refneq{SigmaSSp}
can be represented in the following form 
\[
\varSigma_{jj'}^{(S',S)}=\sum_{\pr R}D_{[0][0]}^{[\lambda']}(\pr R)D_{[0][0]}^{[\lambda]}(\pr R)\sum_{\pr P}\delta_{\lambda_{1},\pr Pj}\delta_{\lambda_{1},\pr R\pr Pj'},
\]
where $\lambda=[N/2+S,N/2-S]$ and $\lambda'=[N/2+S',N/2-S']$.

For $j=j'$, there are $(N-1)!$ permutations $\pr P$ such that $\pr Pj=\lambda_{1}$.
Then
\begin{equation}
\varSigma_{jj}^{(S',S)}=(N-1)!\sum_{\pr R}D_{[0][0]}^{[\lambda']}(\pr R)D_{[0][0]}^{[\lambda]}(\pr R)\delta_{\lambda_{1},\pr R\lambda_{1}}\label{SigmaSSpjj}
\end{equation}
is independent of $j$.

For $j\neq j'$, we have
\begin{multline*}
\sum_{\pr P}\delta_{\lambda_{1},\pr Pj}\delta_{\lambda_{1},\pr R\pr Pj'}=\sum_{l\neq\lambda_{1}}\delta_{\lambda_{1},\pr Rl}\sum_{\pr P}\delta_{l,\pr Pj'}\delta_{\lambda_{1},\pr Pj}\\
=(N-2)!\sum_{l}\delta_{\lambda_{1},\pr Rl}(1-\delta_{\lambda_{1}l})=(N-2)!(1-\delta_{\lambda_{1},\pr R\lambda_{1}}).
\end{multline*}
Then
\begin{multline}
\varSigma_{jj'}^{(S',S)}=\sum_{\pr R}D_{[0][0]}^{[\lambda']}(\pr R)D_{[0][0]}^{[\lambda]}(\pr R)(N-2)!(1-\delta_{\lambda_{1},\pr R\lambda_{1}})\\
=\frac{N!(N-2)!}{f_{S}}\delta_{\lambda\lambda'}-\frac{1}{N-1}\varSigma_{jj}^{(S',S)},\label{SigmaSSpjjp}
\end{multline}
where the last transformation uses Eqs. \refneq{OrthRel} and \refneq{SigmaSSpjj}.
The last expression in \refneq{SigmaSSpjjp} is independent of $j$
and $j'$ and equivalent to \refneq{Sigmajjp}. 

The Young orthogonal matrix elements in \refneq{SigmaSSpjj} have
been calculated by Goddard \cite{goddard1967} in the following way.
Each permutation $\pr R$ can be represented as
\[
\pr R=\prod_{k=1}^{n_{ex}}\pr P_{i'_{k}i''_{k}}\pr P'\pr P'',
\]
where $\pr P'$ are permutations of symbols in the first row of the
Young tableau $[0]$ ($\lambda_{1}$ first symbols), $\pr P''$ are
permutations of symbols in the second row ($\lambda_{2}$ last symbols),
and $\pr P_{i'_{k}i''_{k}}$ transpose symbols between the rows as
$i'_{k}\leq\lambda_{1}$ and $i''_{k}>\lambda_{1}$. Then \cite{goddard1967}
\[
D_{[0][0]}^{[\lambda]}(\pr R)=(-1)^{n_{ex}}\binom{\lambda_{1}}{n_{ex}}^{-1}=(-1)^{n_{ex}}\frac{n_{ex}!(\lambda_{1}-n_{ex})!}{\lambda_{1}!}.
\]

Due to the Kronecker symbols in \refeq{SigmaSSpjj}, the permutations
$\pr P'$ do not affect $\lambda_{1}$ and $i'_{k}\leq\lambda_{1}-1$.
Therefore there are $(\lambda_{1}-1)!$ permutations $\pr P'$ , $\lambda_{2}!$
permutations $\pr P''$, and number of distinct choices of the sets
of $i'_{k}$ and $i''_{k}$ are given by the binomial coefficients
$\binom{\lambda_{1}-1}{n_{ex}}$ and $\binom{\lambda_{2}}{n_{ex}}$,
respectively. Then for $S=S'$ \refeq{SigmaSSpjj} can be transformed
as follows,
\begin{multline*}
\varSigma_{jj}^{(S,S)}=(N-1)!\sum_{n_{ex}=0}^{\lambda_{2}}(\lambda_{1}-1)!\lambda_{2}!\binom{\lambda_{1}-1}{n_{ex}}\binom{\lambda_{2}}{n_{ex}}\binom{\lambda_{1}}{n_{ex}}^{-2}\\
=\frac{(N-1)!(\lambda_{2}!)^{2}}{\lambda_{1}^{2}}\sum_{n_{ex}=0}^{\lambda_{2}}\frac{(\lambda_{1}-n_{ex})!}{(\lambda_{2}-n_{ex})!}(\lambda_{1}-n_{ex}).
\end{multline*}
The sum over $n_{ex}$ can be calculated, leading to \refneq{SigmaSS}
.

If $S'=S-1$ we have
\begin{multline*}
\varSigma_{jj}^{(S-1,S)}=(N-1)!\sum_{n_{ex}=0}^{\lambda_{2}}(\lambda_{1}-1)!\lambda_{2}!\binom{\lambda_{1}-1}{n_{ex}}\binom{\lambda_{2}}{n_{ex}}\binom{\lambda_{1}}{n_{ex}}^{-1}\\
\times\binom{\lambda_{1}-1}{n_{ex}}^{-1}=\frac{(N-1)!(\lambda_{2}!)^{2}}{\lambda_{1}}\sum_{n_{ex}=0}^{\lambda_{2}}\frac{(\lambda_{1}-n_{ex})!}{(\lambda_{2}-n_{ex})!},
\end{multline*}
 giving \refneq{SigmaSm1S} .
\begin{acknowledgments}
The author gratefully acknowledges useful conversations with N. Davidson,
V. Fleurov, I. G. Kaplan, and E. Sela. 
\end{acknowledgments}


\begin{thebibliography}{36}%
\makeatletter
\providecommand \@ifxundefined [1]{%
 \@ifx{#1\undefined}
}%
\providecommand \@ifnum [1]{%
 \ifnum #1\expandafter \@firstoftwo
 \else \expandafter \@secondoftwo
 \fi
}%
\providecommand \@ifx [1]{%
 \ifx #1\expandafter \@firstoftwo
 \else \expandafter \@secondoftwo
 \fi
}%
\providecommand \natexlab [1]{#1}%
\providecommand \enquote  [1]{``#1''}%
\providecommand \bibnamefont  [1]{#1}%
\providecommand \bibfnamefont [1]{#1}%
\providecommand \citenamefont [1]{#1}%
\providecommand \href@noop [0]{\@secondoftwo}%
\providecommand \href [0]{\begingroup \@sanitize@url \@href}%
\providecommand \@href[1]{\@@startlink{#1}\@@href}%
\providecommand \@@href[1]{\endgroup#1\@@endlink}%
\providecommand \@sanitize@url [0]{\catcode `\\12\catcode `\$12\catcode
  `\&12\catcode `\#12\catcode `\^12\catcode `\_12\catcode `\%12\relax}%
\providecommand \@@startlink[1]{}%
\providecommand \@@endlink[0]{}%
\providecommand \url  [0]{\begingroup\@sanitize@url \@url }%
\providecommand \@url [1]{\endgroup\@href {#1}{\urlprefix }}%
\providecommand \urlprefix  [0]{URL }%
\providecommand \Eprint [0]{\href }%
\providecommand \doibase [0]{http://dx.doi.org/}%
\providecommand \selectlanguage [0]{\@gobble}%
\providecommand \bibinfo  [0]{\@secondoftwo}%
\providecommand \bibfield  [0]{\@secondoftwo}%
\providecommand \translation [1]{[#1]}%
\providecommand \BibitemOpen [0]{}%
\providecommand \bibitemStop [0]{}%
\providecommand \bibitemNoStop [0]{.\EOS\space}%
\providecommand \EOS [0]{\spacefactor3000\relax}%
\providecommand \BibitemShut  [1]{\csname bibitem#1\endcsname}%
\let\auto@bib@innerbib\@empty
%</preamble>
\bibitem [{\citenamefont {Bethe}\ and\ \citenamefont {Salpeter}(2008)}]{bethe}%
  \BibitemOpen
  \bibfield  {author} {\bibinfo {author} {\bibfnamefont {H.}~\bibnamefont
  {Bethe}}\ and\ \bibinfo {author} {\bibfnamefont {E.}~\bibnamefont
  {Salpeter}},\ }\href@noop {} {\emph {\bibinfo {title} {Quantum Mechanics of
  One- And Two-Electron Atoms}}}\ (\bibinfo  {publisher} {Dover},\ \bibinfo
  {address} {Mineola, N.Y.},\ \bibinfo {year} {2008})\BibitemShut {NoStop}%
\bibitem [{\citenamefont {Pitaevskii}\ and\ \citenamefont
  {Stringari}(2003)}]{pitaevskii}%
  \BibitemOpen
  \bibfield  {author} {\bibinfo {author} {\bibfnamefont {L.}~\bibnamefont
  {Pitaevskii}}\ and\ \bibinfo {author} {\bibfnamefont {S.}~\bibnamefont
  {Stringari}},\ }\href@noop {} {\emph {\bibinfo {title} {{B}ose-{E}instein
  Condensation}}}\ (\bibinfo  {publisher} {University Press},\ \bibinfo
  {address} {Oxford},\ \bibinfo {year} {2003})\BibitemShut {NoStop}%
\bibitem [{\citenamefont {Landau}\ and\ \citenamefont
  {Lifshitz}(1977)}]{landau}%
  \BibitemOpen
  \bibfield  {author} {\bibinfo {author} {\bibfnamefont {L.}~\bibnamefont
  {Landau}}\ and\ \bibinfo {author} {\bibfnamefont {E.}~\bibnamefont
  {Lifshitz}},\ }\href@noop {} {\emph {\bibinfo {title} {Quantum Mechanics:
  Non-Relativistic Theory}}}\ (\bibinfo  {publisher} {Pergamon Press},\
  \bibinfo {address} {New York},\ \bibinfo {year} {1977})\BibitemShut {NoStop}%
\bibitem [{\citenamefont {Hamermesh}(1989)}]{hamermesh}%
  \BibitemOpen
  \bibfield  {author} {\bibinfo {author} {\bibfnamefont {M.}~\bibnamefont
  {Hamermesh}},\ }\href@noop {} {\emph {\bibinfo {title} {Group Theory and Its
  Application to Physical Problems}}}\ (\bibinfo  {publisher} {Dover},\
  \bibinfo {address} {Mineola, N.Y.},\ \bibinfo {year} {1989})\BibitemShut
  {NoStop}%
\bibitem [{\citenamefont {Elliott}\ and\ \citenamefont
  {Dawber}(1979)}]{elliott}%
  \BibitemOpen
  \bibfield  {author} {\bibinfo {author} {\bibfnamefont {J.}~\bibnamefont
  {Elliott}}\ and\ \bibinfo {author} {\bibfnamefont {P.}~\bibnamefont
  {Dawber}},\ }\href@noop {} {\emph {\bibinfo {title} {Symmetry in physics}}}\
  (\bibinfo  {publisher} {University Press},\ \bibinfo {address} {Oxford},\
  \bibinfo {year} {1979})\BibitemShut {NoStop}%
\bibitem [{\citenamefont {Kaplan}(1975)}]{kaplan}%
  \BibitemOpen
  \bibfield  {author} {\bibinfo {author} {\bibfnamefont {I.}~\bibnamefont
  {Kaplan}},\ }\href@noop {} {\emph {\bibinfo {title} {Symmetry of
  many-electron systems}}}\ (\bibinfo  {publisher} {Academic Press},\ \bibinfo
  {address} {New York},\ \bibinfo {year} {1975})\BibitemShut {NoStop}%
\bibitem [{\citenamefont {Pauncz}(1995)}]{pauncz_symmetric}%
  \BibitemOpen
  \bibfield  {author} {\bibinfo {author} {\bibfnamefont {R.}~\bibnamefont
  {Pauncz}},\ }\href@noop {} {\emph {\bibinfo {title} {The Symmetric Group in
  Quantum Chemistry}}}\ (\bibinfo  {publisher} {CRC Press},\ \bibinfo {address}
  {Boca Raton},\ \bibinfo {year} {1995})\BibitemShut {NoStop}%
\bibitem [{\citenamefont {Heitler}(1927)}]{heitler1927}%
  \BibitemOpen
  \bibfield  {author} {\bibinfo {author} {\bibfnamefont {W.}~\bibnamefont
  {Heitler}},\ }\href@noop {} {\bibfield  {journal} {\bibinfo  {journal} {Z.
  Phys.}\ }\textbf {\bibinfo {volume} {46}},\ \bibinfo {pages} {47} (\bibinfo
  {year} {1927})}\BibitemShut {NoStop}%
\bibitem [{\citenamefont {Myatt}\ \emph {et~al.}(1997)\citenamefont {Myatt},
  \citenamefont {Burt}, \citenamefont {Ghrist}, \citenamefont {Cornell},\ and\
  \citenamefont {Wieman}}]{myatt1997}%
  \BibitemOpen
  \bibfield  {author} {\bibinfo {author} {\bibfnamefont {C.~J.}\ \bibnamefont
  {Myatt}}, \bibinfo {author} {\bibfnamefont {E.~A.}\ \bibnamefont {Burt}},
  \bibinfo {author} {\bibfnamefont {R.~W.}\ \bibnamefont {Ghrist}}, \bibinfo
  {author} {\bibfnamefont {E.~A.}\ \bibnamefont {Cornell}}, \ and\ \bibinfo
  {author} {\bibfnamefont {C.~E.}\ \bibnamefont {Wieman}},\ }\href {\doibase
  10.1103/PhysRevLett.78.586} {\bibfield  {journal} {\bibinfo  {journal} {Phys.
  Rev. Lett.}\ }\textbf {\bibinfo {volume} {78}},\ \bibinfo {pages} {586}
  (\bibinfo {year} {1997})}\BibitemShut {NoStop}%
\bibitem [{\citenamefont {Stamper-Kurn}\ \emph {et~al.}(1998)\citenamefont
  {Stamper-Kurn}, \citenamefont {Andrews}, \citenamefont {Chikkatur},
  \citenamefont {Inouye}, \citenamefont {Miesner}, \citenamefont {Stenger},\
  and\ \citenamefont {Ketterle}}]{stamper1998}%
  \BibitemOpen
  \bibfield  {author} {\bibinfo {author} {\bibfnamefont {D.~M.}\ \bibnamefont
  {Stamper-Kurn}}, \bibinfo {author} {\bibfnamefont {M.~R.}\ \bibnamefont
  {Andrews}}, \bibinfo {author} {\bibfnamefont {A.~P.}\ \bibnamefont
  {Chikkatur}}, \bibinfo {author} {\bibfnamefont {S.}~\bibnamefont {Inouye}},
  \bibinfo {author} {\bibfnamefont {H.-J.}\ \bibnamefont {Miesner}}, \bibinfo
  {author} {\bibfnamefont {J.}~\bibnamefont {Stenger}}, \ and\ \bibinfo
  {author} {\bibfnamefont {W.}~\bibnamefont {Ketterle}},\ }\href {\doibase
  10.1103/PhysRevLett.80.2027} {\bibfield  {journal} {\bibinfo  {journal}
  {Phys. Rev. Lett.}\ }\textbf {\bibinfo {volume} {80}},\ \bibinfo {pages}
  {2027} (\bibinfo {year} {1998})}\BibitemShut {NoStop}%
\bibitem [{\citenamefont {Ho}(1998)}]{ho1998}%
  \BibitemOpen
  \bibfield  {author} {\bibinfo {author} {\bibfnamefont {T.-L.}\ \bibnamefont
  {Ho}},\ }\href {\doibase 10.1103/PhysRevLett.81.742} {\bibfield  {journal}
  {\bibinfo  {journal} {Phys. Rev. Lett.}\ }\textbf {\bibinfo {volume} {81}},\
  \bibinfo {pages} {742} (\bibinfo {year} {1998})}\BibitemShut {NoStop}%
\bibitem [{\citenamefont {Ohmi}\ and\ \citenamefont
  {Machida}(1998)}]{ohmi1998}%
  \BibitemOpen
  \bibfield  {author} {\bibinfo {author} {\bibfnamefont {T.}~\bibnamefont
  {Ohmi}}\ and\ \bibinfo {author} {\bibfnamefont {K.}~\bibnamefont {Machida}},\
  }\href {\doibase 10.1143/JPSJ.67.1822} {\bibfield  {journal} {\bibinfo
  {journal} {J. Phys. Soc. Jpn.}\ }\textbf {\bibinfo {volume} {67}},\ \bibinfo
  {pages} {1822} (\bibinfo {year} {1998})}\BibitemShut {NoStop}%
\bibitem [{\citenamefont {Stamper-Kurn}\ and\ \citenamefont
  {Ueda}(2013)}]{stamper2013}%
  \BibitemOpen
  \bibfield  {author} {\bibinfo {author} {\bibfnamefont {D.~M.}\ \bibnamefont
  {Stamper-Kurn}}\ and\ \bibinfo {author} {\bibfnamefont {M.}~\bibnamefont
  {Ueda}},\ }\href {\doibase 10.1103/RevModPhys.85.1191} {\bibfield  {journal}
  {\bibinfo  {journal} {Rev. Mod. Phys.}\ }\textbf {\bibinfo {volume} {85}},\
  \bibinfo {pages} {1191} (\bibinfo {year} {2013})}\BibitemShut {NoStop}%
\bibitem [{\citenamefont {Guan}\ \emph {et~al.}(2013)\citenamefont {Guan},
  \citenamefont {Batchelor},\ and\ \citenamefont {Lee}}]{guan2013}%
  \BibitemOpen
  \bibfield  {author} {\bibinfo {author} {\bibfnamefont {X.-W.}\ \bibnamefont
  {Guan}}, \bibinfo {author} {\bibfnamefont {M.~T.}\ \bibnamefont {Batchelor}},
  \ and\ \bibinfo {author} {\bibfnamefont {C.}~\bibnamefont {Lee}},\ }\href
  {\doibase 10.1103/RevModPhys.85.1633} {\bibfield  {journal} {\bibinfo
  {journal} {Rev. Mod. Phys.}\ }\textbf {\bibinfo {volume} {85}},\ \bibinfo
  {pages} {1633} (\bibinfo {year} {2013})}\BibitemShut {NoStop}%
\bibitem [{\citenamefont {Yang}(1967)}]{yang1967}%
  \BibitemOpen
  \bibfield  {author} {\bibinfo {author} {\bibfnamefont {C.~N.}\ \bibnamefont
  {Yang}},\ }\href {\doibase 10.1103/PhysRevLett.19.1312} {\bibfield  {journal}
  {\bibinfo  {journal} {Phys. Rev. Lett.}\ }\textbf {\bibinfo {volume} {19}},\
  \bibinfo {pages} {1312} (\bibinfo {year} {1967})}\BibitemShut {NoStop}%
\bibitem [{\citenamefont {Sutherland}(1968)}]{sutherland1968}%
  \BibitemOpen
  \bibfield  {author} {\bibinfo {author} {\bibfnamefont {B.}~\bibnamefont
  {Sutherland}},\ }\href {\doibase 10.1103/PhysRevLett.20.98} {\bibfield
  {journal} {\bibinfo  {journal} {Phys. Rev. Lett.}\ }\textbf {\bibinfo
  {volume} {20}},\ \bibinfo {pages} {98} (\bibinfo {year} {1968})}\BibitemShut
  {NoStop}%
\bibitem [{\citenamefont {Yurovsky}(2014)}]{yurovsky2014}%
  \BibitemOpen
  \bibfield  {author} {\bibinfo {author} {\bibfnamefont {V.~A.}\ \bibnamefont
  {Yurovsky}},\ }\href {\doibase 10.1103/PhysRevLett.113.200406} {\bibfield
  {journal} {\bibinfo  {journal} {Phys. Rev. Lett.}\ }\textbf {\bibinfo
  {volume} {113}},\ \bibinfo {pages} {200406} (\bibinfo {year}
  {2014})}\BibitemShut {NoStop}%
\bibitem [{\citenamefont {Honerkamp}\ and\ \citenamefont
  {Hofstetter}(2004)}]{honerkamp2004}%
  \BibitemOpen
  \bibfield  {author} {\bibinfo {author} {\bibfnamefont {C.}~\bibnamefont
  {Honerkamp}}\ and\ \bibinfo {author} {\bibfnamefont {W.}~\bibnamefont
  {Hofstetter}},\ }\href {\doibase 10.1103/PhysRevLett.92.170403} {\bibfield
  {journal} {\bibinfo  {journal} {Phys. Rev. Lett.}\ }\textbf {\bibinfo
  {volume} {92}},\ \bibinfo {pages} {170403} (\bibinfo {year}
  {2004})}\BibitemShut {NoStop}%
\bibitem [{\citenamefont {Gorshkov}\ \emph {et~al.}(2010)\citenamefont
  {Gorshkov}, \citenamefont {Hermele}, \citenamefont {Gurarie}, \citenamefont
  {Xu}, \citenamefont {Julienne}, \citenamefont {Ye}, \citenamefont {Zoller},
  \citenamefont {Demler}, \citenamefont {Lukin},\ and\ \citenamefont
  {Rey}}]{gorshkov2010}%
  \BibitemOpen
  \bibfield  {author} {\bibinfo {author} {\bibfnamefont {A.~V.}\ \bibnamefont
  {Gorshkov}}, \bibinfo {author} {\bibfnamefont {M.}~\bibnamefont {Hermele}},
  \bibinfo {author} {\bibfnamefont {V.}~\bibnamefont {Gurarie}}, \bibinfo
  {author} {\bibfnamefont {C.}~\bibnamefont {Xu}}, \bibinfo {author}
  {\bibfnamefont {P.~S.}\ \bibnamefont {Julienne}}, \bibinfo {author}
  {\bibfnamefont {J.}~\bibnamefont {Ye}}, \bibinfo {author} {\bibfnamefont
  {P.}~\bibnamefont {Zoller}}, \bibinfo {author} {\bibfnamefont
  {E.}~\bibnamefont {Demler}}, \bibinfo {author} {\bibfnamefont {M.~D.}\
  \bibnamefont {Lukin}}, \ and\ \bibinfo {author} {\bibfnamefont {A.~M.}\
  \bibnamefont {Rey}},\ }\href {\doibase 10.1038/NPHYS1535} {\bibfield
  {journal} {\bibinfo  {journal} {Nat. Phys.}\ }\textbf {\bibinfo {volume}
  {6}},\ \bibinfo {pages} {289} (\bibinfo {year} {2010})}\BibitemShut {NoStop}%
\bibitem [{\citenamefont {Cazalilla}\ \emph {et~al.}(2009)\citenamefont
  {Cazalilla}, \citenamefont {Ho},\ and\ \citenamefont {Ueda}}]{cazalilla2009}%
  \BibitemOpen
  \bibfield  {author} {\bibinfo {author} {\bibfnamefont {M.~A.}\ \bibnamefont
  {Cazalilla}}, \bibinfo {author} {\bibfnamefont {A.~F.}\ \bibnamefont {Ho}}, \
  and\ \bibinfo {author} {\bibfnamefont {M.}~\bibnamefont {Ueda}},\ }\href@noop
  {} {\bibfield  {journal} {\bibinfo  {journal} {New J. Phys.}\ }\textbf
  {\bibinfo {volume} {11}},\ \bibinfo {pages} {103033} (\bibinfo {year}
  {2009})}\BibitemShut {NoStop}%
\bibitem [{\citenamefont {Zhang}\ \emph {et~al.}(2014)\citenamefont {Zhang},
  \citenamefont {Bishof}, \citenamefont {Bromley}, \citenamefont {Kraus},
  \citenamefont {Safronova}, \citenamefont {Zoller}, \citenamefont {Rey},\ and\
  \citenamefont {Ye}}]{zhang2014}%
  \BibitemOpen
  \bibfield  {author} {\bibinfo {author} {\bibfnamefont {X.}~\bibnamefont
  {Zhang}}, \bibinfo {author} {\bibfnamefont {M.}~\bibnamefont {Bishof}},
  \bibinfo {author} {\bibfnamefont {S.~L.}\ \bibnamefont {Bromley}}, \bibinfo
  {author} {\bibfnamefont {C.~V.}\ \bibnamefont {Kraus}}, \bibinfo {author}
  {\bibfnamefont {M.~S.}\ \bibnamefont {Safronova}}, \bibinfo {author}
  {\bibfnamefont {P.}~\bibnamefont {Zoller}}, \bibinfo {author} {\bibfnamefont
  {A.~M.}\ \bibnamefont {Rey}}, \ and\ \bibinfo {author} {\bibfnamefont
  {J.}~\bibnamefont {Ye}},\ }\href {\doibase 10.1126/science.1254978}
  {\bibfield  {journal} {\bibinfo  {journal} {Science}\ }\textbf {\bibinfo
  {volume} {345}},\ \bibinfo {pages} {1467} (\bibinfo {year}
  {2014})}\BibitemShut {NoStop}%
\bibitem [{\citenamefont {Scazza}\ \emph {et~al.}(2014)\citenamefont {Scazza},
  \citenamefont {Hofrichter}, \citenamefont {H\"ofer}, \citenamefont
  {De~Groot}, \citenamefont {Bloch},\ and\ \citenamefont
  {F\"olling}}]{scazza2014}%
  \BibitemOpen
  \bibfield  {author} {\bibinfo {author} {\bibfnamefont {F.}~\bibnamefont
  {Scazza}}, \bibinfo {author} {\bibfnamefont {C.}~\bibnamefont {Hofrichter}},
  \bibinfo {author} {\bibfnamefont {M.}~\bibnamefont {H\"ofer}}, \bibinfo
  {author} {\bibfnamefont {P.}~\bibnamefont {De~Groot}}, \bibinfo {author}
  {\bibfnamefont {I.}~\bibnamefont {Bloch}}, \ and\ \bibinfo {author}
  {\bibfnamefont {S.}~\bibnamefont {F\"olling}},\ }\href@noop {} {\bibfield
  {journal} {\bibinfo  {journal} {Nat. Phys.}\ }\textbf {\bibinfo {volume}
  {10}},\ \bibinfo {pages} {779} (\bibinfo {year} {2014})}\BibitemShut
  {NoStop}%
\bibitem [{\citenamefont {Lieb}\ and\ \citenamefont {Mattis}(1962)}]{lieb1962}%
  \BibitemOpen
  \bibfield  {author} {\bibinfo {author} {\bibfnamefont {E.}~\bibnamefont
  {Lieb}}\ and\ \bibinfo {author} {\bibfnamefont {D.}~\bibnamefont {Mattis}},\
  }\href {\doibase 10.1103/PhysRev.125.164} {\bibfield  {journal} {\bibinfo
  {journal} {Phys. Rev.}\ }\textbf {\bibinfo {volume} {125}},\ \bibinfo {pages}
  {164} (\bibinfo {year} {1962})}\BibitemShut {NoStop}%
\bibitem [{\citenamefont {Guan}\ \emph {et~al.}(2009)\citenamefont {Guan},
  \citenamefont {Chen}, \citenamefont {Wang},\ and\ \citenamefont
  {Ma}}]{guan2009}%
  \BibitemOpen
  \bibfield  {author} {\bibinfo {author} {\bibfnamefont {L.}~\bibnamefont
  {Guan}}, \bibinfo {author} {\bibfnamefont {S.}~\bibnamefont {Chen}}, \bibinfo
  {author} {\bibfnamefont {Y.}~\bibnamefont {Wang}}, \ and\ \bibinfo {author}
  {\bibfnamefont {Z.-Q.}\ \bibnamefont {Ma}},\ }\href {\doibase
  10.1103/PhysRevLett.102.160402} {\bibfield  {journal} {\bibinfo  {journal}
  {Phys. Rev. Lett.}\ }\textbf {\bibinfo {volume} {102}},\ \bibinfo {pages}
  {160402} (\bibinfo {year} {2009})}\BibitemShut {NoStop}%
\bibitem [{\citenamefont {Yang}(2009)}]{yang2009}%
  \BibitemOpen
  \bibfield  {author} {\bibinfo {author} {\bibfnamefont {C.~N.}\ \bibnamefont
  {Yang}},\ }\href {http://stacks.iop.org/0256-307X/26/i=12/a=120504}
  {\bibfield  {journal} {\bibinfo  {journal} {Chin. Phys. Lett.}\ }\textbf
  {\bibinfo {volume} {26}},\ \bibinfo {pages} {120504} (\bibinfo {year}
  {2009})}\BibitemShut {NoStop}%
\bibitem [{\citenamefont {Fang}\ \emph {et~al.}(2011)\citenamefont {Fang},
  \citenamefont {Vignolo}, \citenamefont {Gattobigio}, \citenamefont
  {Miniatura},\ and\ \citenamefont {Minguzzi}}]{fang2011}%
  \BibitemOpen
  \bibfield  {author} {\bibinfo {author} {\bibfnamefont {B.}~\bibnamefont
  {Fang}}, \bibinfo {author} {\bibfnamefont {P.}~\bibnamefont {Vignolo}},
  \bibinfo {author} {\bibfnamefont {M.}~\bibnamefont {Gattobigio}}, \bibinfo
  {author} {\bibfnamefont {C.}~\bibnamefont {Miniatura}}, \ and\ \bibinfo
  {author} {\bibfnamefont {A.}~\bibnamefont {Minguzzi}},\ }\href {\doibase
  10.1103/PhysRevA.84.023626} {\bibfield  {journal} {\bibinfo  {journal} {Phys.
  Rev. A}\ }\textbf {\bibinfo {volume} {84}},\ \bibinfo {pages} {023626}
  (\bibinfo {year} {2011})}\BibitemShut {NoStop}%
\bibitem [{\citenamefont {Daily}\ \emph {et~al.}(2012)\citenamefont {Daily},
  \citenamefont {Rakshit},\ and\ \citenamefont {Blume}}]{daily2012}%
  \BibitemOpen
  \bibfield  {author} {\bibinfo {author} {\bibfnamefont {K.~M.}\ \bibnamefont
  {Daily}}, \bibinfo {author} {\bibfnamefont {D.}~\bibnamefont {Rakshit}}, \
  and\ \bibinfo {author} {\bibfnamefont {D.}~\bibnamefont {Blume}},\ }\href
  {\doibase 10.1103/PhysRevLett.109.030401} {\bibfield  {journal} {\bibinfo
  {journal} {Phys. Rev. Lett.}\ }\textbf {\bibinfo {volume} {109}},\ \bibinfo
  {pages} {030401} (\bibinfo {year} {2012})}\BibitemShut {NoStop}%
\bibitem [{\citenamefont {Harshman}(2014)}]{harshman2014}%
  \BibitemOpen
  \bibfield  {author} {\bibinfo {author} {\bibfnamefont {N.~L.}\ \bibnamefont
  {Harshman}},\ }\href {\doibase 10.1103/PhysRevA.89.033633} {\bibfield
  {journal} {\bibinfo  {journal} {Phys. Rev. A}\ }\textbf {\bibinfo {volume}
  {89}},\ \bibinfo {pages} {033633} (\bibinfo {year} {2014})}\BibitemShut
  {NoStop}%
\bibitem [{\citenamefont {Harshman}(2015)}]{harshman2015}%
  \BibitemOpen
  \bibfield  {author} {\bibinfo {author} {\bibfnamefont {N.~L.}\ \bibnamefont
  {Harshman}},\ }\href@noop {} {\enquote {\bibinfo {title} {One-dimensional
  traps, two-body interactions, few-body symmetries},}\ } (\bibinfo {year}
  {2015}),\ \Eprint {http://arxiv.org/abs/1501.00215} {arXiv:1501.00215}
  \BibitemShut {NoStop}%
\bibitem [{\citenamefont {Yurovsky}(2013)}]{yurovsky2013}%
  \BibitemOpen
  \bibfield  {author} {\bibinfo {author} {\bibfnamefont {V.~A.}\ \bibnamefont
  {Yurovsky}},\ }\href {\doibase 10.1002/qua.24337} {\bibfield  {journal}
  {\bibinfo  {journal} {Int. J. Quantum Chem.}\ }\textbf {\bibinfo {volume}
  {113}},\ \bibinfo {pages} {1436} (\bibinfo {year} {2013})}\BibitemShut
  {NoStop}%
\bibitem [{\citenamefont {Cook}\ and\ \citenamefont
  {De~Lucia}(1971)}]{cook1971}%
  \BibitemOpen
  \bibfield  {author} {\bibinfo {author} {\bibfnamefont {R.~L.}\ \bibnamefont
  {Cook}}\ and\ \bibinfo {author} {\bibfnamefont {F.~C.}\ \bibnamefont
  {De~Lucia}},\ }\href {\doibase http://dx.doi.org/10.1119/1.1976693}
  {\bibfield  {journal} {\bibinfo  {journal} {Am. J. Phys.}\ }\textbf {\bibinfo
  {volume} {39}},\ \bibinfo {pages} {1433} (\bibinfo {year}
  {1971})}\BibitemShut {NoStop}%
\bibitem [{\citenamefont {Edmonds}(1996)}]{edmonds}%
  \BibitemOpen
  \bibfield  {author} {\bibinfo {author} {\bibfnamefont {A.}~\bibnamefont
  {Edmonds}},\ }\href@noop {} {\emph {\bibinfo {title} {Angular Momentum in
  Quantum Mechanics}}}\ (\bibinfo  {publisher} {University Press},\ \bibinfo
  {address} {Princeton},\ \bibinfo {year} {1996})\BibitemShut {NoStop}%
\bibitem [{\citenamefont {Lassalle}(2008)}]{lassalle2008}%
  \BibitemOpen
  \bibfield  {author} {\bibinfo {author} {\bibfnamefont {M.}~\bibnamefont
  {Lassalle}},\ }\href {\doibase 10.1007/s00208-007-0156-5} {\bibfield
  {journal} {\bibinfo  {journal} {Mathematische Annalen}\ }\textbf {\bibinfo
  {volume} {340}},\ \bibinfo {pages} {383} (\bibinfo {year}
  {2008})}\BibitemShut {NoStop}%
\bibitem [{\citenamefont {Murnaghan}(2005)}]{murnaghan}%
  \BibitemOpen
  \bibfield  {author} {\bibinfo {author} {\bibfnamefont {F.}~\bibnamefont
  {Murnaghan}},\ }\href@noop {} {\emph {\bibinfo {title} {The theory of group
  representations}}}\ (\bibinfo  {publisher} {Dover},\ \bibinfo {address}
  {Mineola, N.Y.},\ \bibinfo {year} {2005})\BibitemShut {NoStop}%
\bibitem [{\citenamefont {Dirac}(1929)}]{dirac1929}%
  \BibitemOpen
  \bibfield  {author} {\bibinfo {author} {\bibfnamefont {P.~A.~M.}\
  \bibnamefont {Dirac}},\ }\href {\doibase 10.1098/rspa.1929.0094} {\bibfield
  {journal} {\bibinfo  {journal} {Proc. R. Soc. A}\ }\textbf {\bibinfo {volume}
  {123}},\ \bibinfo {pages} {714} (\bibinfo {year} {1929})}\BibitemShut
  {NoStop}%
\bibitem [{\citenamefont {Goddard}(1967)}]{goddard1967}%
  \BibitemOpen
  \bibfield  {author} {\bibinfo {author} {\bibfnamefont {W.~A.}\ \bibnamefont
  {Goddard}},\ }\href {\doibase 10.1103/PhysRev.157.73} {\bibfield  {journal}
  {\bibinfo  {journal} {Phys. Rev.}\ }\textbf {\bibinfo {volume} {157}},\
  \bibinfo {pages} {73} (\bibinfo {year} {1967})}\BibitemShut {NoStop}%
\end{thebibliography}
\end{document}